\documentclass[pre,aps,superscriptaddress,twocolumn,%
amsmath,amssymb]{revtex4} 

\usepackage{hyperref}
\usepackage{graphics}
\usepackage{epsfig}
\usepackage{epstopdf}
\usepackage{amsmath}
\usepackage{setspace}
\usepackage{graphicx}
\usepackage{dcolumn}
\usepackage{bm}
\usepackage{color}
\usepackage{latexsym}
\usepackage{textcomp}

\begin{document}

\title{Athermal fluctuations in disordered crystals}

\author{Pappu Acharya}
\email{pappuacharya@tifrh.res.in}

\affiliation{Centre for Interdisciplinary Sciences, Tata Institute of Fundamental Research, Hyderabad 500107, India}

\author{Surajit Sengupta}
\email{surajit@tifrh.res.in}

\affiliation{Centre for Interdisciplinary Sciences, Tata Institute of Fundamental Research, Hyderabad 500107, India}

\author{Bulbul Chakraborty}
\email{bulbul@brandeis.edu}

\affiliation{Martin Fisher School of Physics, Brandeis University, Waltham, MA 02454, USA}

\author{Kabir Ramola}
\email{kramola@tifrh.res.in}

\affiliation{Centre for Interdisciplinary Sciences, Tata Institute of Fundamental Research, Hyderabad 500107, India}

\begin{abstract}
We analyze the fluctuations in particle positions and inter-particle forces in disordered crystals composed of jammed soft particles in the limit of weak disorder. We demonstrate that such athermal systems are fundamentally different from their thermal counterparts, characterized by constrained fluctuations of forces perpendicular to the lattice directions. We develop a disorder perturbation expansion in polydispersity about the crystalline state, which we use to derive exact results to linear order. We show that constrained fluctuations result as a consequence of local force balance conditions, and are characterized by non-Gaussian distributions which we derive exactly. We analytically predict several properties of such systems, including the scaling of the average coordination with polydispersity and packing fraction, which we verify with numerical simulations using soft disks with one-sided harmonic interactions.
\end{abstract}


\maketitle

{\it Introduction:} 
Disorder in solids can originate from various sources including quenched impurities, polydispersity in particle sizes, as well as their random thermal motion \cite{phillips2001crystals,chaikin2000principles,nelson1983reentrant}.
In thermal systems, temperature introduces a natural disorder strength that governs the scale of microscopic fluctuations \cite{kubo1966fluctuation}, and consequently controls macroscopic properties. However, many disordered systems when cooled to low temperatures begin to display marked deviations from thermal behaviour \cite{grigera1999observation}, with temperature playing only a weak role in global properties. 
 Examples of such ``athermal" materials include systems displaying glassy behaviour \cite{berthier2011theoretical,kapteijns2019fast}, and jammed packings of particles \cite{jaeger1996granular}. Jammed packings arise in a variety of natural contexts and have been the subject of intense scrutiny in recent years \cite{cates1998jamming, torquato2010jammed,bi2011jamming,charbonneau2015jamming}. At low temperatures such systems are governed purely by the constraints of mechanical equilibrium, with disorder arising from their many possible arrangements.
Although their properties have been sought to be modeled within thermal frameworks \cite{makse2002testing,tighe2010force}, including with temperature-like quantities such as angoricity \cite{edwards1989theory,blumenfeld2009granular}, constructing a statistical mechanical theory for such materials has remained elusive. While many studies have focused on the statistical properties of jammed soft particles \cite{van2009jamming, henkes2007entropy}, in particular close to the unjamming transition \cite{o2002random,o2003jamming,wyart2005rigidity,goodrich2012finite,ramola2017scaling}, a clear understanding of the nature of the jammed phase and its description within a microscopic framework is still lacking.
 It is therefore important to develop exact theoretical techniques with which to treat such systems.
 

\begin{figure}[t!]
\includegraphics[width=0.9\linewidth]{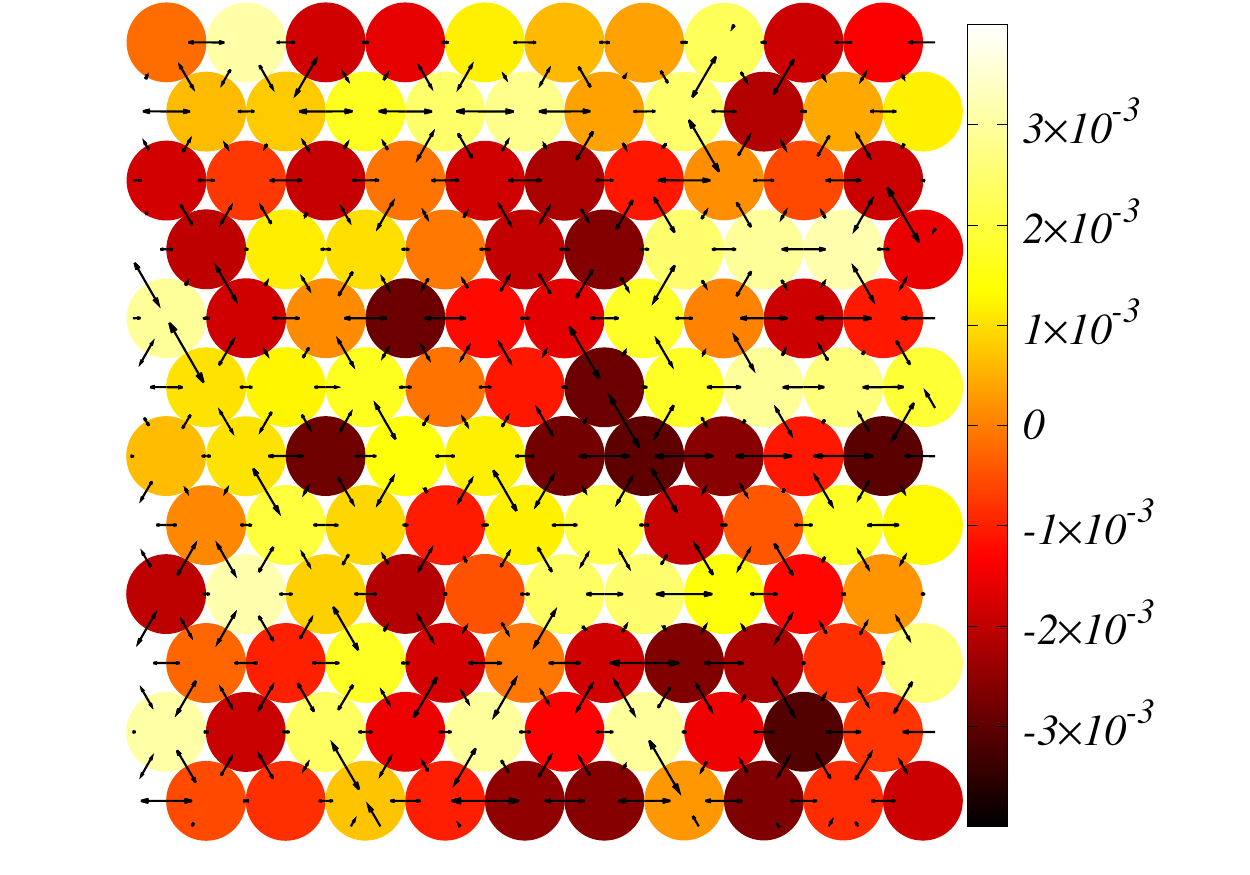}
\caption{A section of a disordered crystal composed of jammed soft particles. The particles are colored according to their incremental size $\Delta \sigma_i = \sigma_i -  \sigma_0$, where $\sigma_i$ are their radii and $\sigma_0 = 1/2$. When $\Delta \sigma_i = 0$ the system settles into a triangular lattice. The black arrows represent the change in the inter-particle forces from their values in the pure crystal in response to the change in radii. For small disorder (polydispersity) the forces fluctuate primarily along the lattice directions.
} 
\label{single_defect_figure}
\end{figure}

In this Letter we present exact results for fluctuations and distribution functions in jammed soft particle packings. We show that athermal disorder characterized by polydispersity, induces fundamentally different statistical properties in jammed systems as compared to thermal disorder. In order to make analytic predictions we make use of a well-known paradigm where exact results are obtainable: that of {\it crystals}. The stability and response of crystals to disorder has been an enduring problem in physics, and
several frameworks have focused on thermal fluctuations in crystals, as well as properties of asperities, disinclinations and defects \cite{phillips2001crystals, nelson1979dislocation}. However the properties of crystals composed of jammed particles, where polydispersity introduces an athermal disorder have been relatively less studied \cite{goodrich2014solids,tong2015crystals,charbonneau2019glassy}.
We demonstrate that in such athermal crystals the constraints of mechanical equilibrium lead to highly constrained fluctuations of the inter-particle forces, in comparison to thermal fluctuations which violate these local constraints. We introduce a disorder perturbation expansion about the crystalline state which allows us to predict several properties of the system including the fluctuations in positions, forces and bond angles. We use this theory to analytically predict non-Gaussian distributions for the components of forces  orthogonal to the original lattice directions, a feature absent from thermal fluctuations.

We consider a system of frictionless disks in two dimensions
interacting through a pairwise one-sided potential that is now paradigmatic in the study of soft particles and deformable foams \cite{durian1995foam,o2002random}.
The interaction is given by
\begin{eqnarray}
\nonumber
V_{\sigma_{ij}}(\vec{r}_{ij}) &=& \frac{\epsilon}{\alpha}\left(1- \frac{| \vec{r}_{ij}|}{\sigma_{ij}}\right)^\alpha ~~\textmd{for}~~ r_{ij} < \sigma_{ij},\\
&=& 0 ~~~~~~~~~~~~~~~~~~~~~\textmd{for}~~r_{ij} \ge \sigma_{ij}.
\label{energy_law}
\end{eqnarray}
Here $\vec{r}_{ij} = \vec{r}_{i} - \vec{r}_{j}$ is the vector distance between the particles $i$ and $j$ located at positions $\vec{r}_i$ and $\vec{r}_j$ respectively, and $\sigma_{i j} = \sigma_i + \sigma_j$ is 
the sum of the radii $\sigma_i$ and $\sigma_j$ of the two particles. Since the interaction potential only depends on the scalar distance $|\vec{ r}_{ij}|$, the system can only sustain normal forces.
In this work we present results for the harmonic case $\alpha = 2$, however our techniques can be generalized to systems with different $\alpha$.
The forces are determined by 
\begin{equation}
\vec{f}_{ij} = \frac{\epsilon}{\sigma_{ij}} \left(1- \frac{| \vec{r}_{ij}|}{\sigma_{ij}}\right)^{\alpha-1} \hat{r}_{ij},
\label{force_law_equation}
\end{equation}
where $\hat{r}_{ij}$ is the unit vector along the  $\vec{r}_{ij}$ direction.
The ground states of the system consist of configurations in mechanical equilibrium, i.e. each particle is in force balance with
\begin{equation}
\sum_{j} f^{x}_{ij} = 0, ~~~\sum_{j} f^{y}_{ij} = 0, ~~~~\forall ~i.
\label{force_balance_eq}
\end{equation}
Here $f^{x(y)}_{ij}$ are the $x(y)$ components of the forces between particles $i$ and $j$, and the sum extends over all particles $j$ in contact with particle $i$.

When all the radii are equal, the minimum energy configuration is a crystalline state with the positions of the centers $\{\vec{r}_{i,0}\}$ forming a triangular lattice (see Fig. \ref{single_defect_figure}). 
The distribution of the forces in the crystalline system is given by
\begin{equation}
p(\vec{f}_{ij}) = \frac{1}{6 f_0}\delta(|f|- f_0)\delta(\theta - \theta_{ij}^{0}),
\label{pure_crystal_force_eq}
\end{equation}
where the magnitude of the force $f_0$ depends on the packing fraction $\phi$, and  $\theta_{ij}^{0}$ is the angle between the particles $i$ and $j$ in the triangular lattice arrangement. Choosing the equal radii to be $\sigma_i = \sigma_0 = 1/2$, the magnitude $f_0$ is given by (see Supplementary Material for details \cite{SI})
\begin{equation}
f_0 = 1- \sqrt{\frac{\phi_c}{\phi}}.
\label{packing_fraction_eq}
\end{equation}
Here $\phi_c$ is the packing fraction of the marginal crystal with no overlaps between particles, with $\phi_c = \pi/\sqrt{12}\approx 0.9069$. 
The force vanishes for the marginal crystal $\Delta \phi = \phi - \phi_{c} = 0$. 

\begin{figure}[t!]
\includegraphics[width=1.1\linewidth]{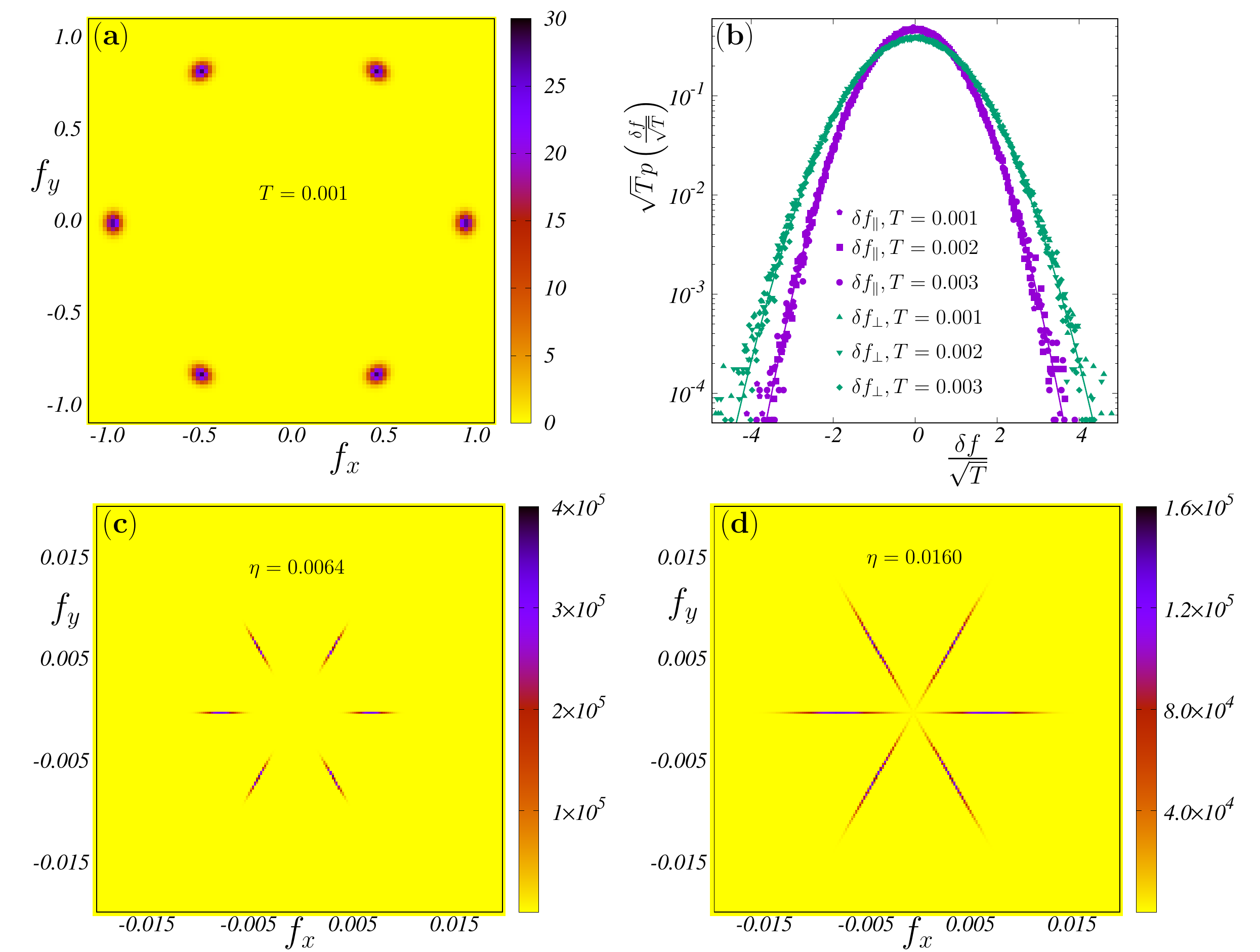}
\caption{ {\bf (a)} Two-dimensional distributions of the forces $p(f_x, f_y)$ in the thermal crystal. {\bf (b)} Scaled distributions of the fluctuations of the components of forces along ($\delta f_{||}$) and orthogonal ($\delta f_{\perp}$) to the original lattice directions. Both display Gaussian fluctuations (marked with solid lines) with a variance proportional to the temperature $T$. {\bf (c)}-{\bf (d)} In contrast $p(f_x, f_y)$ for the athermal system displays highly constrained fluctuations orthogonal to the original lattice directions as the polydispersity ($\eta$) is increased. Here $\phi = 0.92$ and the number of particles is $N = 2500$.
}
\label{con_f_figure}
\end{figure}

{\it Thermal versus athermal fluctuations:}
We begin by analyzing the differences in the force distributions produced by thermal disorder (characterized by a temperature $T$), and athermal disorder (characterized by a polydispersity $\eta$). 
For the thermal case, we perform finite temperature Monte Carlo simulations. We begin at the ground state, by creating a triangular lattice of $N$ equal sized disks in a commensurate rectangular box ($L_y = \frac{\sqrt{3}}{2} L_x$). The fluctuations in the positions are then sampled using the interactions given by Eq. (\ref{energy_law}) at a finite temperature $T$.  
As the temperature is increased from $T = 0$ to a finite value, the distribution of the forces deviate from the pure crystalline delta function peaks in Eq. (\ref{pure_crystal_force_eq}), with a mean $f_0$  and a standard deviation  $\propto \sqrt{T}$. This broadening in the force distribution occurs in the components of the forces along the lattice directions $f_{||}$ as well as  orthogonal to the lattice directions $f_{\perp}$. Both these distributions display Gaussian behaviour as shown in Figs. \ref{con_f_figure} {\bf (a)} and {\bf (b)}.

Similarly, we can characterize the fluctuations in forces in the athermal system with increasing polydispersity.  In this case the temperature is set at $T=0$, and the system samples only the ground state for every realization of the disorder, i.e. states in mechanical equilibrium.
Disorder is introduced into the system by varying the radii of particles. Starting from the state with all radii equal $\sigma_i = \sigma_0 =  \frac{1}{2}$ (i.e. all $\sigma_{ij} = 1$), the radii are incremented as
\begin{equation}
 \sigma_i = \left(1 + \eta~ \xi_i \right) \sigma_0,
 \label{change_radii_law}
\end{equation} 
where $\xi_i$ are independent identically distributed  (i.i.d.)  random variables chosen from an underlying distribution $p(\xi)$. We choose this to be a uniform distribution in the interval $ \left[-\frac{1}{2}, \frac{1}{2}\right]$ \cite{tong2015crystals}.
The polydispersity parameter $\eta$ quantifies the amount of athermal disorder. 
 For each realization of the noise $\{ \xi_i \}$, the system is allowed to settle into a minimum energy configuration as a response to the change in radii. As $\eta$ is increased from zero, 
the forces once again deviate from their pure crystalline values. We measure the distribution of the components of forces parallel to the original lattice directions $p(f_{||})$ as the strength of the disorder is increased. This distribution is well-fit by a Gaussian with the mean $f_0$, and standard deviation $\propto \eta$. This seems to suggest that these fluctuations can be modeled by an effective {\it thermal} Hamiltonian, with the polydispersity playing the role of a temperature $T \propto \eta^2$.  However, a striking difference between thermal and athermal fluctuations emerges when one considers the two dimensional distributions of the forces (as shown in Figs. \ref{con_f_figure} {\bf (c)} and {\bf (d)}). The distribution of the orthogonal components $p(f_{\perp})$ is highly confined with width $\sigma_{\perp} \ll \sigma_{||}$. On this scale the fluctuations perpendicular to the unperturbed lattice directions are negligible in comparision to the fluctuations along the lattice directions. Since the forces in the system are normal, these constrained fluctuations also imply highly constrained fluctuations in the bond angles $\theta_{ij}$.  Moreover, $p(f_{\perp})$ displays significant non-Gaussian behaviour with increasing polydispersity. Remarkably, as we show below, this distribution can be predicted theoretically. The non-Gaussian nature of this distribution, along with the exact prediction is displayed in Fig. \ref{fy_figure}.


{\it Disorder Perturbation Expansion:}
In order to theoretically characterize athermal fluctuations, we analyze the response of the crystalline state in the limit of weak disorder. This allows us to treat the polydispersity as a perturbation about the crystalline state. Here we present an outline of the computation, with details provided in the Supplemental Material \cite{SI}. The radii in Eq. (\ref{change_radii_law}) can be expressed as $\sigma_i = \sigma_0 + \delta \sigma_i$, with $\delta \sigma_i \sim \mathcal{O}(\eta)$.  As a response, the positions of the particles deviate from their crystalline values  $\{x_{i,0}, y_{i,0}\}$ to a new mechanical equilibrium configuration $\{x_{i}, y_{i}\}$. These positions can also be expressed as an expansion in the disorder strength $\eta$, which to lowest order is 
\begin{eqnarray}
\nonumber
x_i &=& x_{i,0} + \delta x_i,\\
y_i &=& y_{i,0} + \delta y_i.
\end{eqnarray}
Here $\delta x_i$ and $\delta y_i$ are small perturbations of $\mathcal{O}(\eta)$.
The forces in Eq. (\ref{force_law_equation}) can then be expressed in terms of these variables as an expansion, which to linear order is given by
\begin{eqnarray}
\nonumber
\delta {f}_{ij}^{x} &=& C_{ij}^{xx}\delta x_{ij} + C_{ij}^{xy} \delta y_{ij} + C_{ij}^{x \sigma} \delta \sigma_{ij},\\
\delta {f}_{ij}^{y} &=& C_{ij}^{yx}\delta x_{ij} + C_{ij}^{yy} \delta y_{ij} + C_{ij}^{y \sigma} \delta \sigma_{ij}.
\label{linearized_force_eq}
\end{eqnarray}
Here $\delta x_{ij} = \delta x_{i} -\delta x_{j}$, $\delta y_{ij} = \delta y_{i} -\delta y_{j}$ whereas $\delta \sigma_{ij} = \delta \sigma_{i} + \delta \sigma_{j}$. The coefficients $C_{ij}^{\alpha \beta}$ can be expressed purely in terms of the positions of the crystalline state, and are translationally invariant. 
We can exploit this invariance by considering the equations of mechanical equilibrium (Eq. (\ref{force_balance_eq})) in Fourier space. Using the forces in Eq. (\ref{linearized_force_eq}), the equations for force balance can be expressed in Fourier space as
\begin{equation}
\left(
\begin{matrix} 
A^{xx}(\vec{k}) & A^{xy}(\vec{k}) \\
A^{yx}(\vec{k}) & A^{yy}(\vec{k})
\end{matrix}
\right)
\left(
\begin{matrix} 
\delta x (\vec{k}) \\
\delta y (\vec{k})
\end{matrix}
\right)
= 
\delta \sigma (\vec{k})
\left(
\begin{matrix} 
D^{x} (\vec{k}) \\
D^{y} (\vec{k})
\end{matrix}
\right).
\label{fourier_matrix_eq}
\end{equation}
Here $\vec{k} \equiv (k_x,k_y) = \left( \frac{2 \pi l}{2 N}, \frac{2 \pi m}{N} \right)$ are the reciprocal lattice vectors of the triangular lattice. The above equation can be interpreted as the change in the position fields in response to the ``charges" introduced by the variation in the particle radii. The inversion of this equation in Fourier space yields
\begin{equation}
\delta x(\vec{k}) = \alpha(\vec{k}) \delta \sigma (\vec{k}); ~~\delta y(\vec{k}) = \beta(\vec{k}) \delta \sigma (\vec{k}).
\label{deltax_k_eq}
\end{equation}
The exact expressions for $\alpha(\vec{k})$ and $\beta(\vec{k})$ are rather cumbersome and we provide a detailed derivation in the Supplemental Material \cite{SI}.

\begin{figure}[t!]
\includegraphics[width=1.1\linewidth]{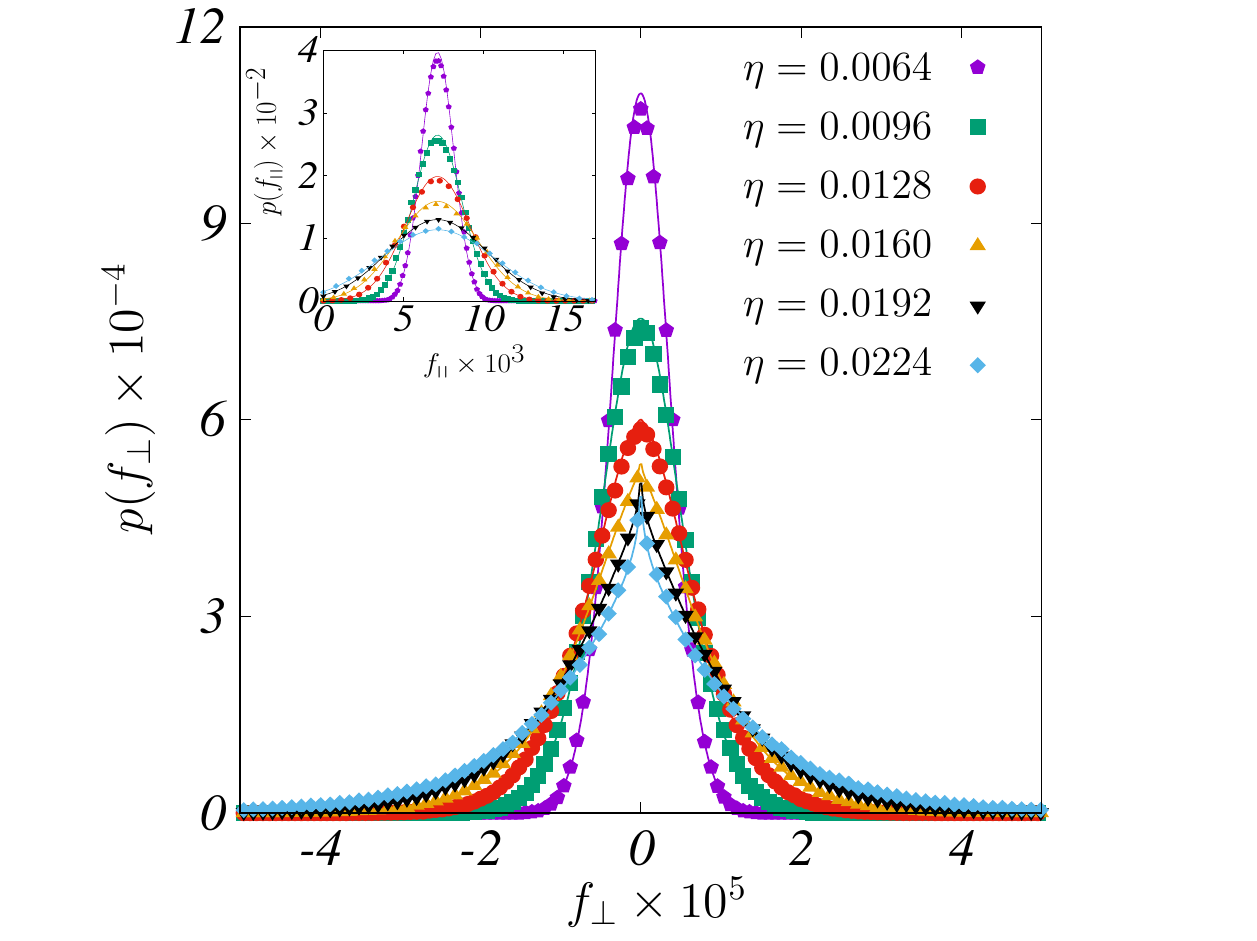}
\caption{The distribution of $f_{\perp}$ in the athermal system, displaying near-perfect agreement with the theoretical prediction in Eq. (\ref{perpendicular_distribution}) (solid lines). Here $\phi = 0.92$ and $N = 2500$. This distribution exhibits pronounced non-Gaussian behaviour as the disorder strength is increased, in contrast to the thermal system in Fig. \ref{con_f_figure}{\bf (b)}. {\bf (Inset)} The distribution of $f_{||}$ displays Gaussian behaviour. The predictions from the theory are displayed with solid lines.} 
\label{fy_figure}
\end{figure} 

We are now in a position to derive the fluctuations in the positions of the particles in response to the athermal disorder. For example, the fluctuations in $x$ are given by
\begin{equation}
\langle \delta x_{i}^2 \rangle = \frac{1}{2L^2} \sum_{\vec{k}} \alpha(\vec{k})  \alpha(-\vec{k}) \langle \delta \sigma^2\rangle,
\end{equation}
where $N = L^2$ is the number of particles in the system. The fluctuations in the radii are i.i.d. variables with $\langle \delta \sigma^2\rangle = \eta^2/48$. We note that this expression provides the {\it exact} leading order coefficient of the variance in the positions. In the Supplemental Material \cite{SI} we show the excellent agreement between the above theoretical prediction and our numerical simulations.


{\it Non-Gaussian force distributions:}
One of the surprising characteristics of athermal fluctuations in disordered crystals is the appearance of non-Gaussian probability distributions in the components of the forces perpendicular to the lattice directions. Remarkably, these distributions can be derived analytically using the perturbation theory in polydispersity as we show below.
The fluctuations in the force magnitudes $|f| = \sqrt{f_{||}^2 + f_{\perp}^2} \approx f_{||}$ can be obtained from the position fluctuations using Eq. (\ref{linearized_force_eq}) (see Supplemental Material for details \cite{SI}). As the inversion in Fourier space expresses the forces in the system as a linear combination of the fluctuations in the radii, the distribution of $|f|$ can be shown to be a Gaussian with mean $f_0$ and standard deviation  $0.157 ~\eta$.  The distribution of $f_{||}$ for various polydispersities is shown in the inset of Fig. \ref{fy_figure}, along with the theoretically predicted Gaussian distributions showing excellent agreement.
Following a similar argument as for $f_{||}$, the fluctuations in the positions can also be used to derive the fluctuations in the bond angles $\delta \theta_{ij} = \theta_{ij} -\theta_{ij}^{0}$. The distribution of $\sin(\delta \theta)$ is once again a Gaussian distribution with mean $0$, and standard deviation  $0.0813 ~\eta$.  At linear order, the correlations between these variables is small in comparison to their individual fluctuations, and we may treat them as uncorrelated (see Supplemental Material \cite{SI}).
These distributions can then be used to derive the distribution of $f_{\perp} = |f| \sin{(\delta \theta)}$.
Since the product of two Gaussian variables with non-zero means exhibits non-Gaussian behavior \cite{craig1936frequency}, we find that the distribution of $f_{\perp}$ indeed begins to deviate from a Gaussian distribution at large polydispersities. This distribution is given by (with $s \equiv \sin \delta \theta$) 
\begin{equation}
p(f_{\perp}) =\int_{0}^{\infty}  d |f|  \int_{-1}^{1} d s ~p(|f|) p(s) \delta \left( f_{\perp} - |f|  s \right),
\label{perpendicular_distribution}
\end{equation}
 and can be evaluated analytically \cite{cui2016exact} (see Supplemental Material \cite{SI}).
The pronounced non-Gaussian behaviour of the distribution $p(f_{\perp})$ computed using the above expression is displayed in Fig. \ref{fy_figure}, showing near-perfect agreement with distributions obtained from direct numerical simulations.

\begin{figure}[t!]
\includegraphics[width=1.1\linewidth]{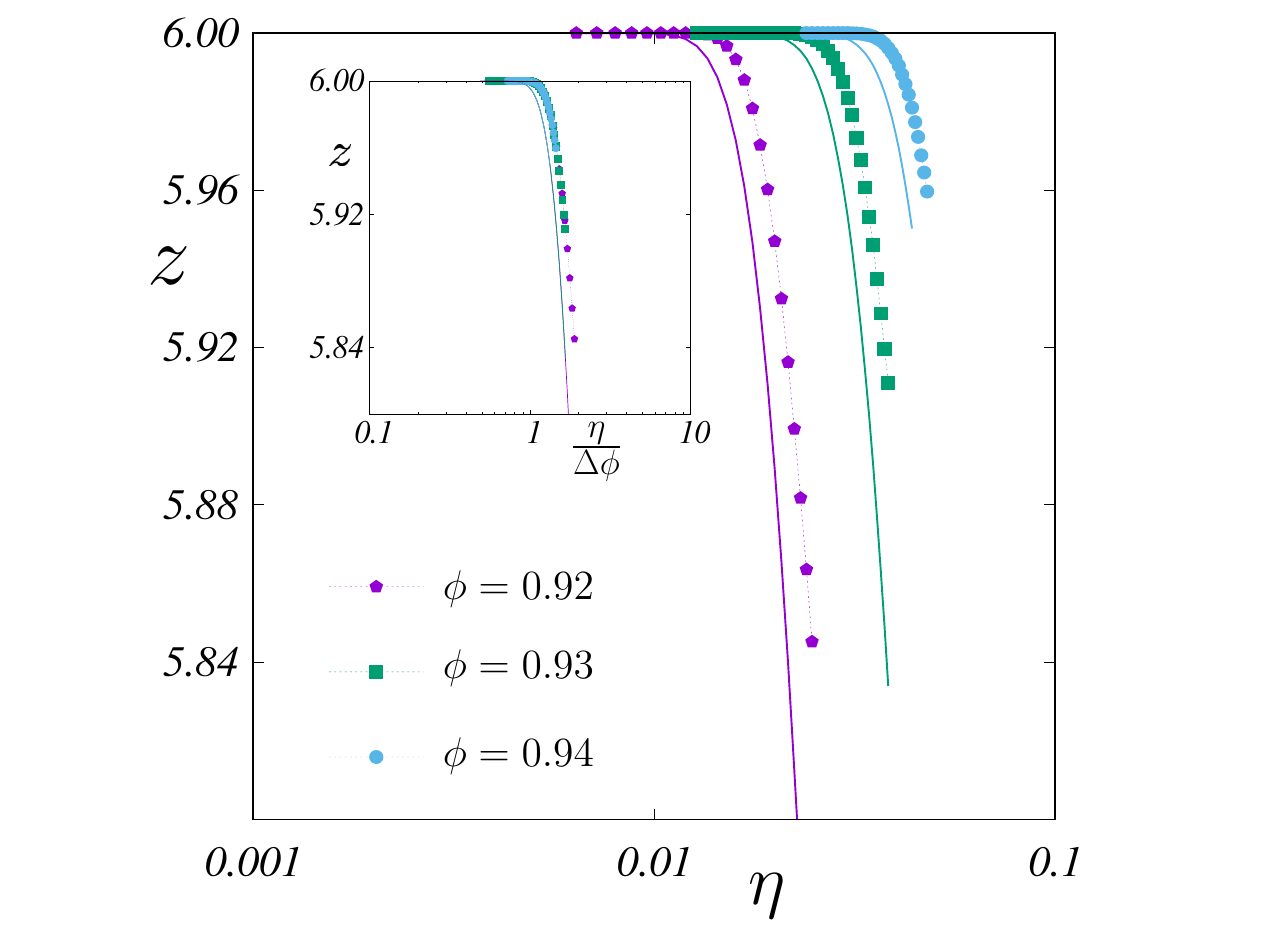}
\caption{Variation of the average coordination with the strength of the disorder for different packing fractions ($\phi$). The points represent data from simulations, and the solid lines represent the theoretical prediction in Eq. (\ref{coordination_eq}).  {\bf (Inset)} The scaling collapse of the average coordination with the predicted scaling variable $\frac{\eta}{\Delta\phi}$. Here $N = 2500$.
} 
\label{zz_figure}
\end{figure}


{\it Average Coordination:}
Finally, we use the microscopic predictions from our theory to compute a macroscopic property of the system, namely the dependence of the average coordination on other global parameters such as the polydispersity and packing fraction. 
Since the magnitude of the forces in the system can only take positive values, the negative regions in the theoretical distribution of $p(|f|)$ represent the broken contacts in the system.  Consequently the average coordination to lowest order in $\eta$ is given by $z = 6 \int_{0}^\infty p(|f|)d|f|$. However, we have shown that the distribution of  $|f| \approx f_{||}$ is a Gaussian with mean $f_0$, and standard deviation $0.157 ~\eta$.
The dependence of the mean value of the force $f_0$ on the packing fraction can be obtained by Taylor expanding Eq. (\ref{packing_fraction_eq}) upto first order in $\Delta \phi$, we have $f_0 = \Delta\phi/2 \phi_c$.  
This yields a theoretical prediction for the average coordination in the system
\begin{equation}
z = 3 \left(1 + \textrm{erf}\left(\mathcal{C} \Delta\phi/\eta \right)\right),
\label{coordination_eq}
\end{equation}
with $\mathcal{C}^{-1} = 0.4440~\phi_c$.
Since all coordination related quantities can be obtained from the underlying force distribution $p(|f|)$, this theory predicts that the average coordination as well as the susceptibilities for different packing fractions can be collapsed with the scaling variable $\frac{\eta}{\Delta\phi}$, as has been observed numerically in previous studies \cite{tong2015crystals}. We plot the variation of the average coordination with polydispersity along with the above theoretical prediction in Fig. \ref{zz_figure}. Once again this theory does well in tracking the behaviour of this non-trivial global parameter, and indeed predicts the scaling with  $\frac{\eta}{\Delta\phi}$ perfectly. However, we note that the numerical values of $z$ obtained from simulations display a small deviation from the predicted theoretical curve. We attribute this to the system spanning rearrangements induced by contact breaking events which cannot be exactly modeled within a linear framework. In the Supplemental Material \cite{SI} we provide details of this non-linear contact breaking process observed in the simulations.

{\it Discussion:}
 In this Letter we have presented exact results for the fluctuations of particle positions and inter-particle forces in jammed soft particle crystals. The limit of small disorder allowed us to express the local force balance conditions as a set of linear equations relating the particle coordinates and the particle radii. Exploiting the crystal periodicity of the original lattice, the leading order coefficients of the fluctuations of positions, forces and relative bond angles could be analytically predicted. This allowed us to express the distribution of the components of the forces perpendicular to the lattice directions as a product of two Gaussian variables, which displays non-Gaussian fluctuations.
 
Since analytic results are rare in the study of disordered jammed matter, it is surprising that many properties of disordered crystals are amenable to theoretical computation. At higher disorder strengths, enough bonds break in the system, and the angular fluctuations become deconfined, which could be considered to be a non-linear effect. Indeed this system exhibits a non-trivial phase transition to a disordered amorphous phase with increasing disorder \cite{tong2015crystals}.
This transition is characterized by diverging fluctuations in coordination numbers over different realizations, and it would be interesting to understand this behaviour by studying interactions between defects in the near-crystalline system.
Finally, it would also be interesting to use the techniques developed in this paper to predict how microscopic constraints of force balance in such athermal materials give rise to an emergent elasticity at large length scales \cite{nampoothiri2020emergent}, with non-trivial stress transmission and rheological properties.
  
  {\it Acknowledgments:}
  We thank Smarajit Karmakar and Srikanth Sastry for useful discussions. The work of BC has been supported by NSF-CBET Grant No. 1916877 and 1605428 as well as BSF Grant No. 2016118. This project was funded by intramural funds at TIFR Hyderabad from the Department of Atomic Energy (DAE).

\bibliography{constrained_fluctuations_bibliography.bib} 


\clearpage

\begin{widetext}

\begin{appendix}
 

\section*{\large Supplemental Material for ``Athermal fluctuations in disordered crystals"}

In this document we provide supplemental figures and details related to the results presented in the main text.

\maketitle




\subsection{Simulating thermal disorder}
 
In this section we study thermal crystals using finite temperature Monte Carlo simulations. We simulate two distinct systems (i) a crystal of equal sized particles governed by a force law (Eq. (\ref{force_law_equation}) in the main text), and (ii) a crystal with forces not related to the inter-particle distances, but force balanced at all times. In the latter case we posit a quadratic Hamiltonian governing the forces that couples to the temperature of the system. In both cases we observe Gaussian fluctuations of the forces along all directions, showing that the constrained fluctuations observed in athermal systems arise from both local force balance conditions {\it and} the force law relating the positions to the forces in the system.

\begin{figure}[ht!]
\includegraphics[width=1.05\linewidth]{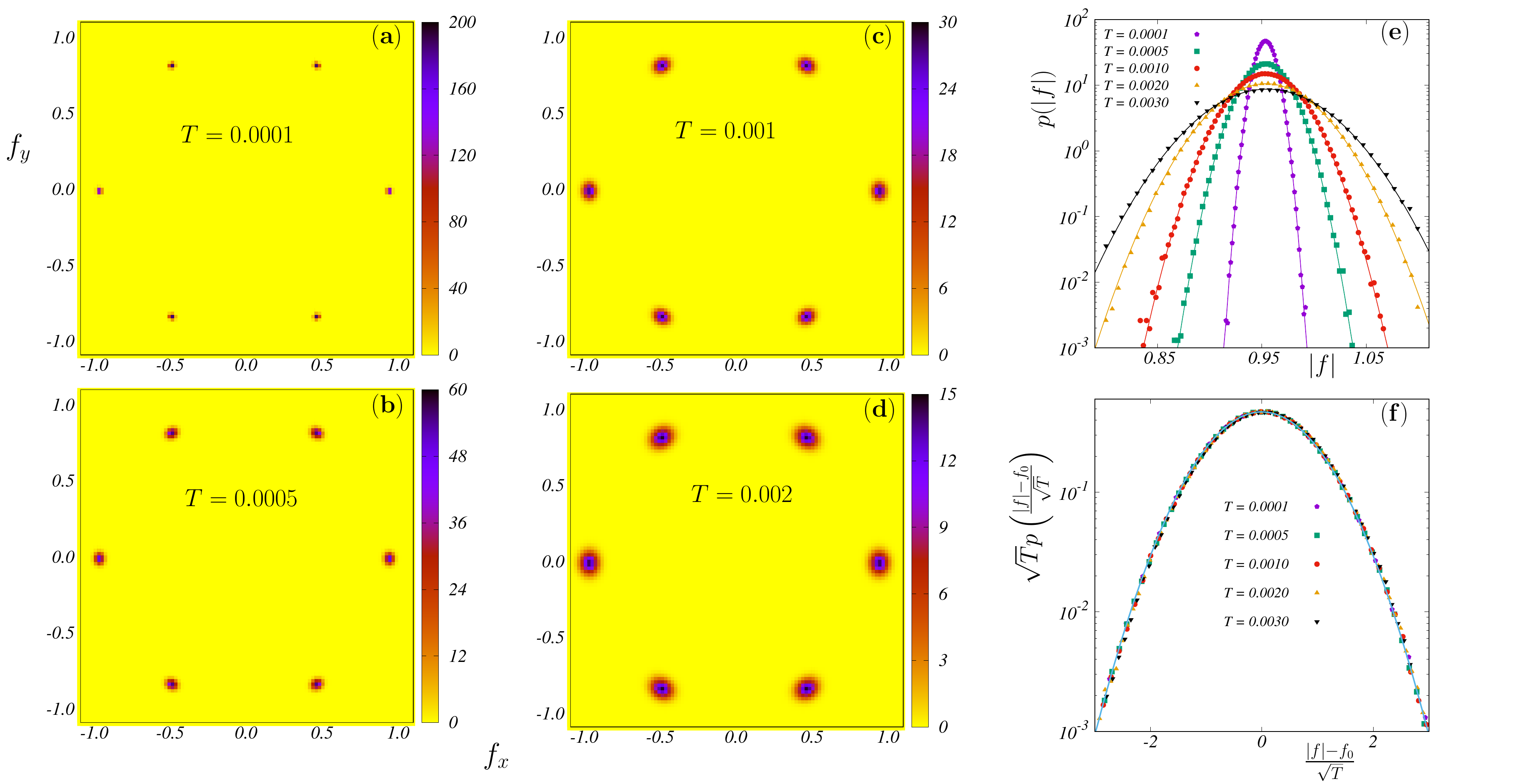}
\caption{{\bf(a)} - {\bf (d)} Two dimensional distributions $p(f_x,f_y)$ of the forces $\vec{f} \equiv (f_x,f_y)$ in the system at different temperatures obtained from Monte Carlo simulations of thermal crystals. The fluctuations about the crystalline values are Gaussian with a width $\propto \sqrt{T}$, exhibiting thermal broadening along the lattice directions as well as directions perpendicular to the lattice. {\bf (e)} The distributions of the magnitudes of the forces $p(|f|)$ are Gaussian at all simulated temperatures, the bold lines represent best-fit Gaussians. {\bf (f)} These distributions can be collapsed with the single scaling variable $\zeta = (|f| - f_0)/\sqrt{T}$.} 
\label{mc_main_figure}
\end{figure}

\subsubsection{Thermal Crystal}

In our simulations of thermal crystals, the interactions between particles is modeled with the one-sided harmonic law given in Eq. (\ref{energy_law}) in the main text. All particles have equal radii with $\sigma_i = \sigma_0 = \frac{1}{2}$, forming a triangular lattice at $T=0$. At finite temperatures, we allow fluctuations in the particle positions of magnitude $\propto \sqrt{T}$, and use Metropolis sampling to accept or reject configurations.
It should be noted that the states sampled by these simulations violate the local force balance conditions at every time, since the system is not at an energy minimum.
The two dimensional distributions of the forces obtained from these simulations is plotted in Fig. \ref{mc_main_figure} {\bf (a)} - {\bf (d)}. At $T=0$ the distribution is peaked at six locations governed by the crystalline angles with magnitude $f_0$ (Eq. (\ref{pure_crystal_force_eq}) in the main text). These peaks spread as the temperature increases, exhibiting Gaussian fluctuations in both $f_{\perp}$ and $f_{||}$ (shown in Fig. \ref{con_f_figure} {\bf (b)} in the main text). Consequently the distribution of the magnitude of the forces $|f|$ also displays Gaussian fluctuations as displayed in Fig. \ref{mc_main_figure} {\bf (e)}. These distributions can be collapsed with the scaling variable $\zeta = (|f| - f_0)/\sqrt{T}$ as shown in Fig. \ref{mc_main_figure} {\bf (f)}.

\subsubsection{``Thermal'' force balanced crystal}

\begin{figure}[h!]
\includegraphics[width=1.05
\linewidth]{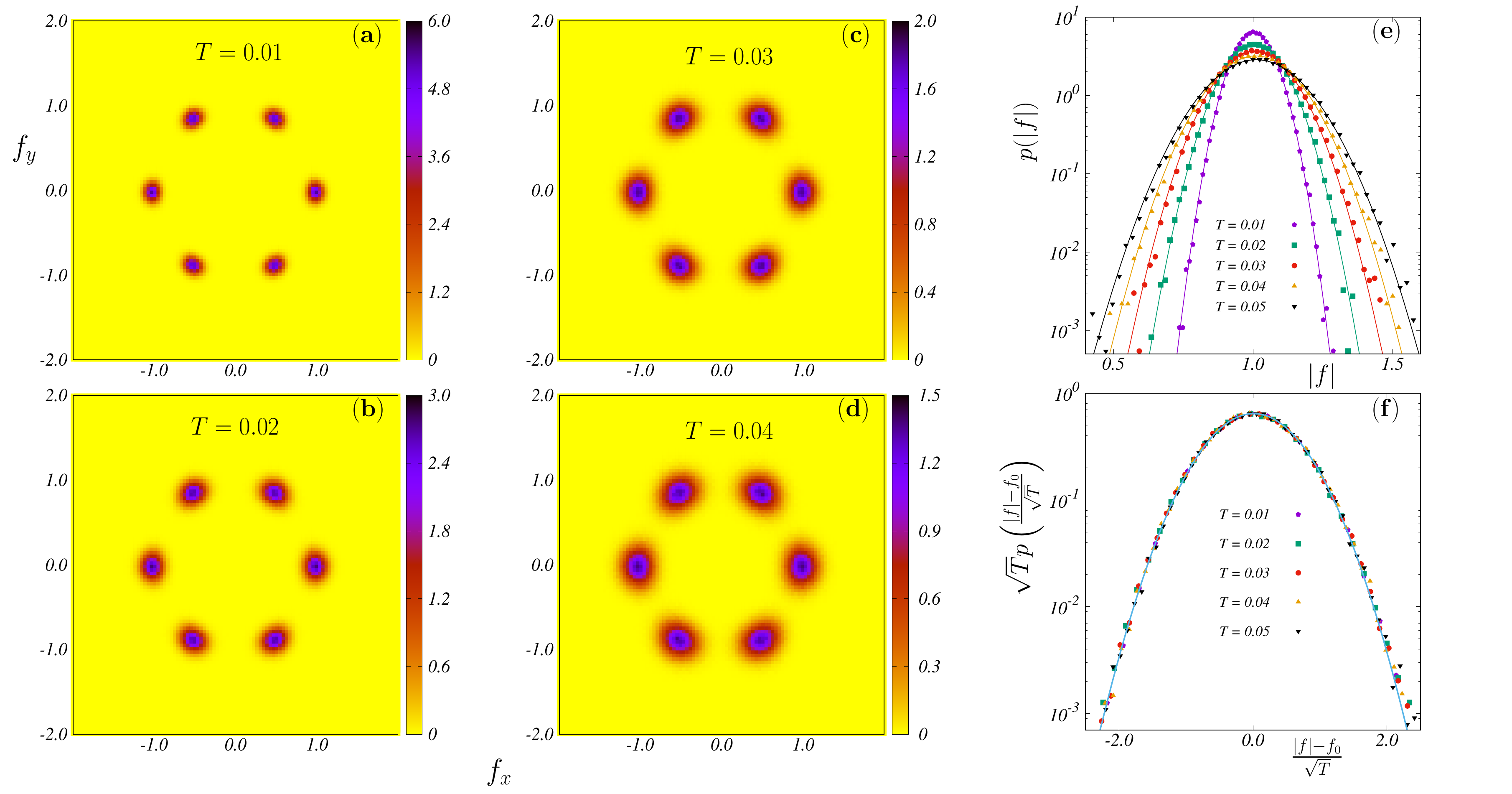}
\caption{{\bf (a)} - {\bf (d)} Two dimensional distributions $p(f_x,f_y)$ of the forces $\vec{f} \equiv (f_x,f_y)$ in the system at different temperatures obtained from Monte Carlo simulations of crystals in force balance without a force law. The fluctuations about the crystalline values are Gaussian with a width $\propto \sqrt{T}$, once again exhibiting thermal broadening along both the lattice directions as well as directions perpendicular to the lattice. {\bf (e)} The distributions of the magnitudes of the forces $p(|f|)$ are Gaussian at all simulated temperatures, the bold lines represent best-fit Gaussians. {\bf (f)} These distributions can be collapsed with the single scaling variable $\zeta = (|f| - f_0)/\sqrt{T}$.} 
\label{mc_fb_figure}
\end{figure}

Since constrained fluctuations arising in athermal systems originate from local force balance conditions, it is interesting to ask whether such fluctuations can be obtained from an effective Hamiltonian {\it with force balance} on every particle.  We therefore postulate an effective Hamiltonian of the harmonic form
\begin{equation}
H = \epsilon_f \sum_{\langle i j \rangle} \left(\vec{f}_{ij}-\vec{f}^0_{ij}\right)^2.
\label{force_hamiltonian}
\end{equation}
Here $\vec{f}^0_{ij}$ are the value of the forces in the pure crystal ($\eta = 0$). In our simulations we set the stiffness $\epsilon_f = 1$. Since we also incorporate the force balance constraint on every grain
\begin{equation}
\sum_j\vec{f_{ij}}=0,
\end{equation}
the finite temperature partition function of such a system is given by
\begin{equation}
Z(\beta) =   \int \prod_{ij} d \vec{f}_{ij}  \prod_i \delta\left(\sum_j \vec{f_{ij}}\right) \exp \left(- \beta H \right),
\end{equation}
where $\beta = 1/T$ is the inverse temperature. We note that since the position degrees of freedom are absent in the above formalism, this represents a system in force balance, but with forces not originating from an underlying force law such as Eq. (\ref{force_law_equation}) in the main text.
In order to incorporate the local force balance constraints, we parameterize the forces in the system in terms of auxiliary fields placed on the voids between grains, termed ``height fields'' \cite{ball2002stress,ramola2017stress}. 
We then perform Monte Carlo simulations by allowing fluctuations in these height fields of magnitude $\propto \sqrt{T}$, and use Metropolis sampling to accept or reject configurations with the Hamiltonian in Eq. (\ref{force_hamiltonian}). Results from these simulations are presented in Fig. \ref{mc_fb_figure}. We find that this system exhibits properties identical to a thermal crystal without force balance (with an underlying force law). We therefore conclude that the constrained fluctuations exhibited by athermal systems originate from the local force balance conditions imposed on the {\it positions} that yield the forces in the system. This is precisely what the disorder perturbation expansion developed in the main text accomplishes.

\subsection{Simulating athermal disorder}

 

Our simulation of athermal disorder follows a standard technique for creating jammed packings of frictionless disks. We begin with a triangular lattice arrangement of the particles, with all radii equal ($\sigma_0 = 1/2$). We then change the particle radii as mentioned in Eq. (\ref{change_radii_law}) in the main text. Finally, we rescale all the radii to keep the packing fraction intact (see Section \ref{boundaryconditions_subsection}).
In order to minimize the energy of the system we use the FIRE (Fast Inertial Relaxation Engine) algorithm \cite{bitzek2006structural}, followed by a molecular dynamics update. FIRE is simple to incorporate and rapidly leads to a minimum energy configuration. In our implementation we compute the power $P = \vec{F}.\vec{v}$ in the entire system at every time step. If $P > 0$, 
the velocity is set to $\vec{v} \rightarrow(1 - \beta)  \vec{v} + \beta \hat{F} |\vec{v}| $, the time step is increased as $\Delta t = \Delta t f_{\textrm{inc}}$ upto a maximum value $\Delta t = \Delta t_{\textrm{max}}$ and $\beta$ is changed to $\beta f_\beta$. However if $P < 0$, the velocity is set to zero, the time step is decreased as $\Delta t = \Delta t f_{\textrm{dec}}$ and $\beta$ is reset back to its initial value $\beta_{\textrm{start}}$. After each such step, we return to the molecular dynamics simulations and update the system with the new velocities. We repeat this process until a desired threshold for force balance in our system is achieved. In our simulations we set $\beta = \beta_{\textrm{start}} = 0.01$, 
$\Delta t = 0.0001$, $\Delta t_{\textrm{max}} =0.001$, $f_{\beta} = 0.99$, $f_{\textrm{inc}} = 1.1$, and $f_{\textrm{dec}} = 0.5$.

\subsubsection{Boundary Conditions}
\label{boundaryconditions_subsection}
In our simulation of athermal disorder, we work in a fixed packing fraction ensemble (similar to Ref. \cite{tong2015crystals}). The athermal perturbation, due to the change in particle radii, changes the packing fraction of the system. Therefore, in our simulations, for a given realization of the disorder (i.e. incremental sizes of the particles), we rescale the radii of all particles (by the same factor) in order to maintain a fixed packing fraction. 
Note that this rescaling does not affect the force distributions to {\it linear order}, as we show below. 
The packing fraction $\phi$ is determined through the equation 
\begin{equation}
\phi = \frac{1}{V}\sum_{i=1}^{N}\pi \sigma_i^2,
\end{equation}
where $V = L_x L_y$ is the volume of the system.
The addition of athermal disorder changes the radii as $\sigma_i \to \sigma_0 (1  + \eta x_i)$, with $x_i$ drawn from a uniform distribution in the interval $\left[ -\frac{1}{2}, \frac{1}{2}\right]$. This leads to the following expansion for the packing fraction
\begin{equation}
\phi' = \phi^0 + \frac{\sum_{i=1}^{N} \eta^2 x_i^2}{V} = \phi^0 + \frac{\eta^2}{48 V}+ \mathcal{O}(\eta^3),
\end{equation}
where $\phi^0$ is the packing fraction of the unperturbed crystal, and since $\sum_{i=1}^{N} \eta x_i \to 0$ for large $N$. Hence, the effect of the rescaling only contributes at order $\eta^2$. 
Similarly, the generalized expression for the force in the pure crystal (Eq. (\ref{generalized_force_eq})) is not affected by the rescaling to leading order.
Finally, we have also checked that the force distributions we obtain do not differ before and after rescaling to linear order.

\subsection{Parallel and Perpendicular Force Components}

\begin{figure}[ht!]
\includegraphics[width=0.3\linewidth]{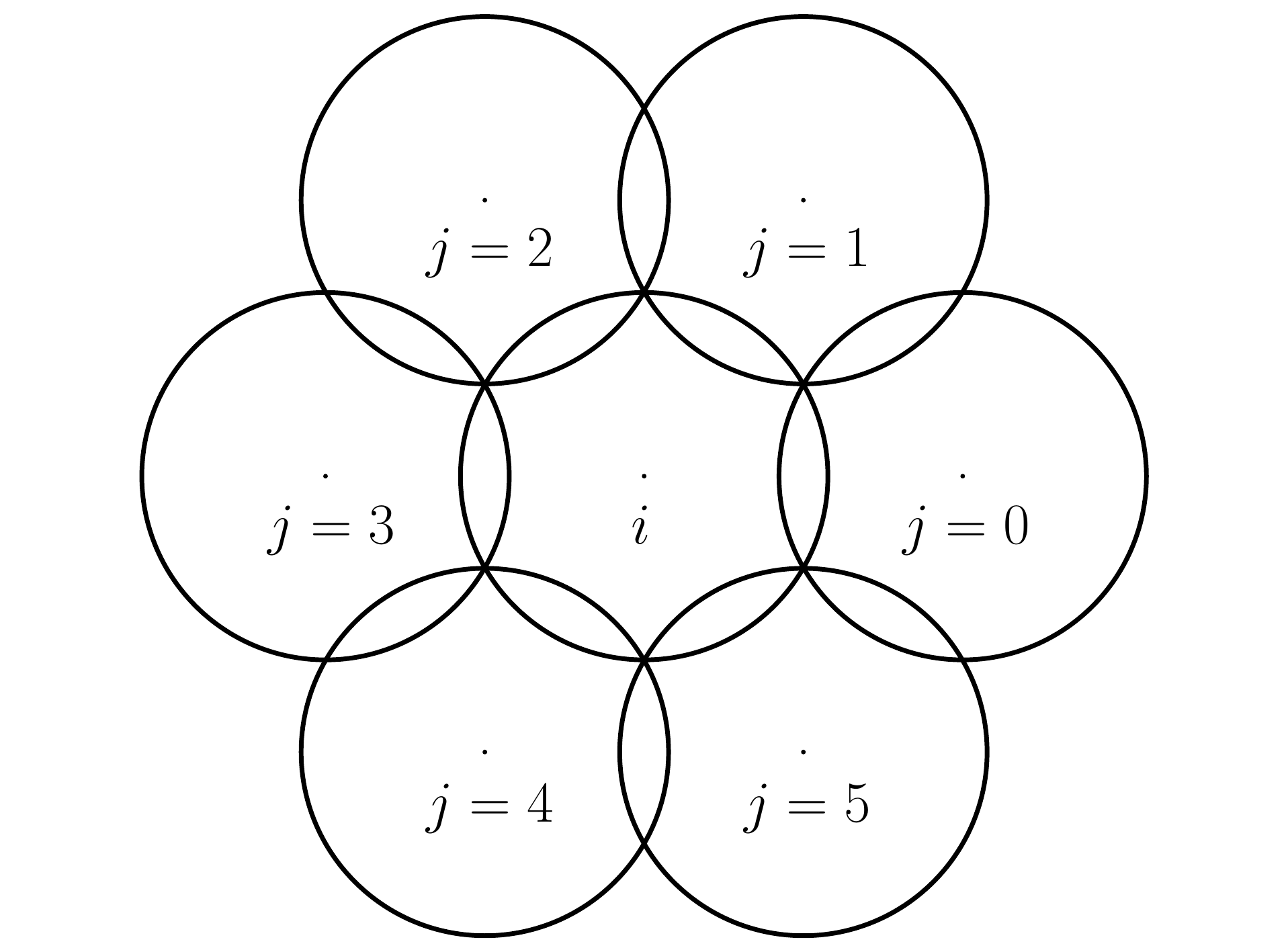}
\caption{The labeling convention. The six neighbours of every particle $i$ are labeled as $j = 0$ to $5$. The bond angles between these particles can take any of six values (depending on $i$ and $j$) with the positive $x$-axis, $\theta_{ij}^{0} = 0$, $\pi/3$, $2 \pi/3$, $\pi$, $4 \pi /3$, and $5 \pi/3$.} 
\label{lattice_figure}
\end{figure}

For a perfect crystal (i.e. no disorder) every particle $i$ has six neighbours $j$.
The bond angles between these particles can take any of six values (depending on $i$ and $j$) with the positive $x$-axis, $\theta_{ij}^{0} = 0$, $\pi/3$, $2 \pi/3$, $\pi$, $4 \pi /3$, and $5 \pi/3$. 
In the main text, we have termed these six directions the `lattice directions'. 
The directions orthogonal to these six directions are termed the ‘perpendicular (or orthogonal) directions’. In the pure crystal, the forces lie precisely along the lattice directions. However, when disorder is introduced, the positions of the particles shift, leading to a finite component along the orthogonal directions. We define these components with respect to the {\it original} lattice directions as
\begin{eqnarray}
\nonumber
f_{\parallel} = |f_{ij}| \cos(\theta_{ij}-\theta_{ij}^{0}),\\
f_{\perp} = |f_{ij}| \sin(\theta_{ij}-\theta_{ij}^{0}).
\end{eqnarray}
Here $|f_{ij}|$ is the magnitude of the force between particles $i$ and $j$ in the disordered ground state. Therefore, in effect we are resolving the perturbed forces along the original crystal structure. Remarkably, this resolution of the forces uncovers a sharp distinction between thermal and athermal crystals, as we show in our study.

\subsection{Forces in the pure crystal} 

In this section we relate the magnitude of the force $f_0$ in a pure crystal ($\eta = 0$) to the packing fraction $\phi$.
Since we set the radius of every particle in the crystalline state to $\sigma_0$, the linear dimensions $L_y = \frac{\sqrt{3}}{2} L_x$ of the system are determined by the packing fraction $\phi$ through the equation
\begin{equation}
\phi = \frac{N \pi \sigma_0^2}{\frac{\sqrt{3}}{2}L_x^2}.
\label{phi_eq}
\end{equation}
For a fixed $L_x$, this leads to the relation
\begin{equation}
\frac{\phi}{\phi_{c}} = \frac{\sigma_0^2}{\sigma_{0,c}^2},
\label{phi_ratio_eq}
\end{equation}
where $\phi_{c}$ and $\sigma_{0,c}$ are the packing fraction and radii of particles in the marginal crystal (with no overlaps between particles) respectively.
As the system is in a triangular lattice arrangement of $\sqrt{N} \times \sqrt{N}$ particles, $L_x$ can be related to the overlap $\Delta r$ between the particles as
\begin{equation}
L_x = \sqrt{N}(2 \sigma_0 - \Delta r).
\end{equation}
 Therefore
 \begin{equation}
\Delta r = 2 \sigma_0 - \frac{L_x}{\sqrt{N}}.
 \label{delta_r_eq}
\end{equation}
Next, we can use this to determine the radii of the particles in the marginal crystal by setting $\Delta r =0$, yielding
 \begin{equation}
 \sigma_{0,c} ^2 = \frac{L_x^2}{4N}.
\label{phi_c_eq}
\end{equation}  
We note that inserting this value into Eq. (\ref{phi_eq}) yields the packing fraction for the hexagonal close packed structure $\phi_c = \frac{\pi}{\sqrt{12}}$. We next relate the overlaps between particles to the inter-particle forces.
Combining Eqs. (\ref{phi_ratio_eq}), (\ref{delta_r_eq}) and (\ref{phi_c_eq}), we have
\begin{equation}
\Delta r = 2 \sigma_0 \left( 1- \sqrt{\frac{\phi_c}{\phi}} \right).
\label{overlap_eq}
\end{equation}
Setting $\epsilon = 1$ in the force law in Eq. (\ref{force_law_equation}) in the main text, we have
\begin{equation}
f_0=\frac{1}{2 \sigma_0} \left( \frac{\Delta r}{2 \sigma_0} \right).
\end{equation}
Using the expression for the overlap in Eq. (\ref{overlap_eq}) in the above expression, we arrive at
\begin{equation}
f_0 =  \frac{1}{2 \sigma_0} \left(1 - \sqrt{\frac{\phi_{c}}{\phi}}\right).
\label{generalized_force_eq}
\end{equation} 
Finally, setting the value $\sigma_0 = \frac{1}{2}$ (as in our simulations), we have
\begin{equation}
f_0 = 1 - \sqrt{\frac{\phi_{c}}{\phi}},
\end{equation}    
which is Eq. (\ref{packing_fraction_eq}) in the main text.


\subsection{Linearized force balance equations} 

In this section we provide details of the disorder perturbation expansion developed in the main text.
The general form of the interaction between particles is given by
\begin{eqnarray}
\nonumber
V_{\sigma_{ij}}(\vec{r}_{ij}) &=& \frac{\epsilon}{\alpha}\left(1- \frac{| \vec{r}_{ij}|}{\sigma_{ij}}\right)^\alpha ~~\textmd{for}~~ r_{ij} < \sigma_{ij},\\
&=& 0 ~~~~~~~~~~~~~~~~~~~~~\textmd{for}~~r_{ij} \ge \sigma_{ij}.
\label{energy_law_supp}
\end{eqnarray}
The forces in the system are determined by the inter-particle distances as
\begin{equation}
\vec{f}_{ij} = \frac{\epsilon}{\sigma_{ij}} \left( 1- \frac{|{r}_{ij}|}{\sigma_{ij}} \right)^{\alpha -1} \hat{r}_{ij}.
\end{equation}
The two components of the forces can be expressed as
\begin{eqnarray}
\nonumber
{f}_{ij}^{x} = \frac{\epsilon}{\sigma_{ij}} \left( 1- \frac{\sqrt{x_{ij}^2 + y_{ij}^2}}{\sigma_{ij}} \right)^{\alpha -1} \frac{x_{ij}}{\sqrt{x_{ij}^2 + y_{ij}^2}},\\
{f}_{ij}^{y} = \frac{\epsilon}{\sigma_{ij}} \left( 1- \frac{\sqrt{x_{ij}^2 + y_{ij}^2}}{\sigma_{ij}} \right)^{\alpha -1} \frac{y_{ij}}{\sqrt{x_{ij}^2 + y_{ij}^2}}.
\label{force_component_eq}
\end{eqnarray}
We note that these equations are non-linear in the components $x_{ij}$ and  $y_{ij}$. 
We next treat the polydispersity as a perturbation, with 
\begin{equation}
\sigma_i = \sigma_0 + \delta \sigma_i.
\end{equation}
As a response to this perturbation from the crystalline state, the positions of the particles also change as
\begin{eqnarray}
\nonumber
x_i &=& x_{i,0} + \delta x_i,\\
y_i &=& y_{i,0} + \delta y_i.
\end{eqnarray}
Expanding Eq. (\ref{force_component_eq}) to linear order in $\delta x_i$ and  $\delta y_i$, the change in the forces due to the disorder can be expressed as
\begin{eqnarray}
\nonumber
\delta {f}_{ij}^{x} &=& C_{ij}^{xx}\delta x_{ij} + C_{ij}^{xy} \delta y_{ij} + C_{ij}^{x \sigma} \delta \sigma_{ij},\\
\delta {f}_{ij}^{y} &=& C_{ij}^{yx}\delta x_{ij} + C_{ij}^{yy} \delta y_{ij} + C_{ij}^{y \sigma} \delta \sigma_{ij},
\label{linearized_expansion}
\end{eqnarray}
where the coefficients $C_{ij}^{ \beta \gamma}(\phi)$ only depend on the packing fraction $\phi$.
These coefficients are translationally invariant, i.e. they do not depend on the particle index $i$. We compute them for a particle $i$, with the neighbouring particles labeled $j = 0$ to $5$ (see Fig. \ref{lattice_figure}).
The coefficients can then be expressed as (setting $\epsilon = 1$)
\begin{eqnarray}
\nonumber
C^{xx}_{ij}(\phi) &=&   -\frac{\left(1-\sqrt{\frac{\phi _c}{\phi }}\right){}^{\alpha -2} \left(\alpha  \sqrt{\frac{\phi _c}{\phi }}+\cos \left(\frac{2 \pi  j}{3}\right) \left((\alpha -2) \sqrt{\frac{\phi _c}{\phi
   }}+1\right)-1\right)}{2 \sqrt{\frac{\phi _c}{\phi }}},\\
\nonumber
C^{xy}_{ij}(\phi) &=&   -\frac{\sin \left(\frac{2 \pi  j}{3}\right) \left(1-\sqrt{\frac{\phi _c}{\phi }}\right){}^{\alpha -2} \left((\alpha -2) \sqrt{\frac{\phi _c}{\phi }}+1\right)}{2 \sqrt{\frac{\phi _c}{\phi }}},\\
\nonumber
C^{x \sigma}_{ij}(\phi) &=&  \cos \left(\frac{\pi  j}{3}\right) \left(1-\sqrt{\frac{\phi _c}{\phi }}\right){}^{\alpha -2} \left(\alpha  \sqrt{\frac{\phi _c}{\phi }}-1\right),\\
\nonumber
C^{yy}_{ij}(\phi) &=&  \frac{\left(1-\sqrt{\frac{\phi _c}{\phi }}\right){}^{\alpha -2} \left(\alpha  \left(-\sqrt{\frac{\phi _c}{\phi }}\right)+\cos \left(\frac{2 \pi  j}{3}\right) \left((\alpha -2) \sqrt{\frac{\phi
   _c}{\phi }}+1\right)+1\right)}{2 \sqrt{\frac{\phi _c}{\phi }}},\\
\nonumber
C^{yx}_{ij}(\phi) &=&  -\frac{\sin \left(\frac{2 \pi  j}{3}\right) \left(1-\sqrt{\frac{\phi _c}{\phi }}\right){}^{\alpha -2} \left((\alpha -2) \sqrt{\frac{\phi _c}{\phi }}+1\right)}{2 \sqrt{\frac{\phi _c}{\phi }}},\\
C^{y \sigma}_{ij}(\phi) &=&  \sin \left(\frac{\pi  j}{3}\right) \left(1-\sqrt{\frac{\phi _c}{\phi }}\right){}^{\alpha -2} \left(\alpha  \sqrt{\frac{\phi _c}{\phi }}-1\right),
\end{eqnarray}
where $\phi_c = \pi/\sqrt{12}$ is the packing fraction of the marginal crystal (with zero overlaps).

\subsubsection{Coefficients for harmonic interactions}
For the harmonic case ($\alpha = 2$) that we study using simulations, the coefficients in the linearized force balance expansion have a particularly simple form, given by
\begin{eqnarray}
\nonumber
C^{xx}_{ij}(\phi) &=&   \frac{\sin ^2\left(\frac{\pi  j}{3}\right)}{\sqrt{\phi_c/\phi}}-1,\\
\nonumber
C^{xy}_{ij}(\phi) &=&   -\frac{\sin \left(\frac{\pi  j}{3}\right) \cos \left(\frac{\pi  j}{3}\right)}{\sqrt{\phi_c/\phi}},\\
\nonumber
C^{x \sigma}_{ij}(\phi) &=&  \left(2-\frac{1}{\sqrt{\phi_c/\phi} }\right) \sqrt{\phi_c/\phi} \cos \left(\frac{\pi  j}{3}\right),\\
\nonumber
C^{yy}_{ij}(\phi) &=&  \frac{\cos ^2\left(\frac{\pi  j}{3}\right)}{\sqrt{\phi_c/\phi}}-1,\\
\nonumber
C^{yx}_{ij}(\phi) &=&   -\frac{\sin \left(\frac{\pi  j}{3}\right) \cos \left(\frac{\pi  j}{3}\right)}{\sqrt{\phi_c/\phi}},\\
C^{y \sigma}_{ij}(\phi) &=&  \left(2-\frac{1}{\sqrt{\phi_c/\phi}}\right) \sqrt{\phi_c/\phi} \sin \left(\frac{\pi  j}{3}\right).
\end{eqnarray}


\subsection{Mean Value of Forces}

In this section we describe the variation in the mean value of the forces, with the introduction of  both thermal and athermal disorder.

\begin{figure*}[t!]
\centering
{\bf (a)}
\includegraphics[scale=0.45]{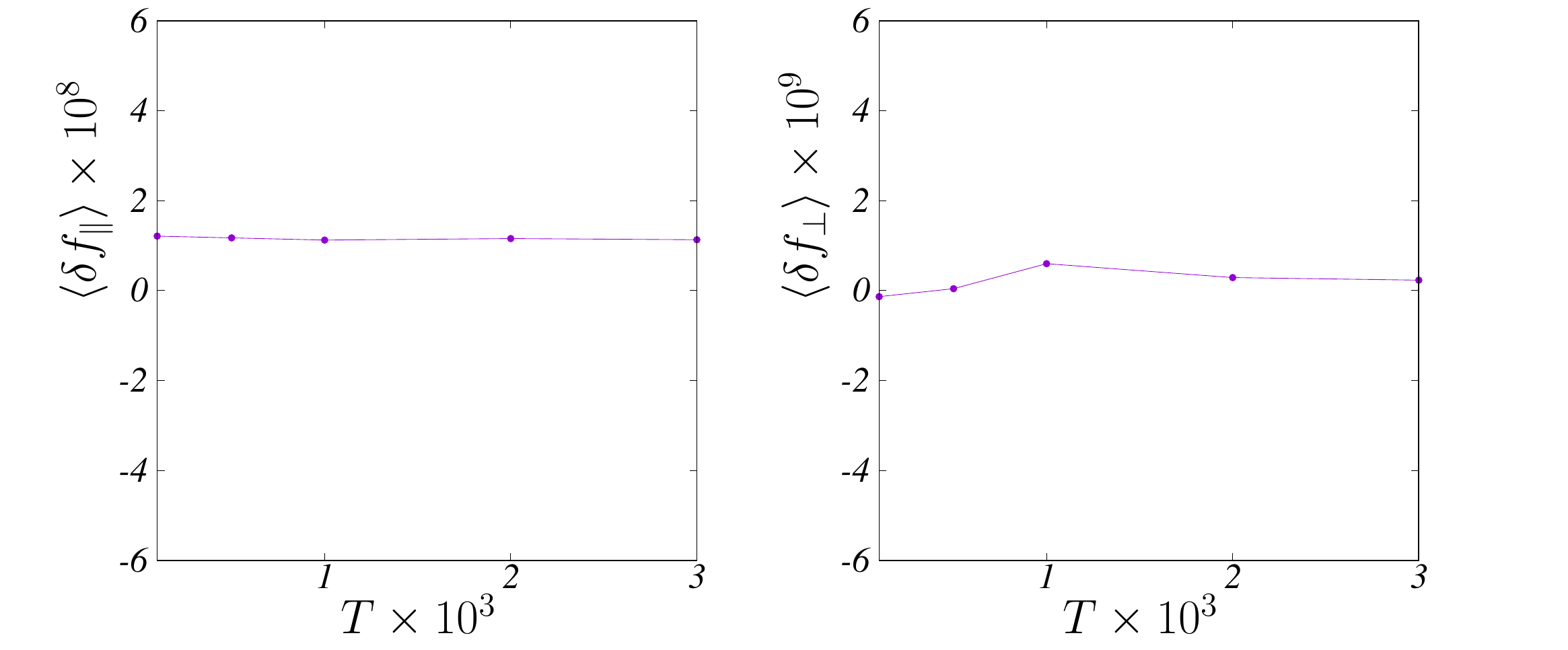}
{\bf (b)}
\includegraphics[scale=0.45]{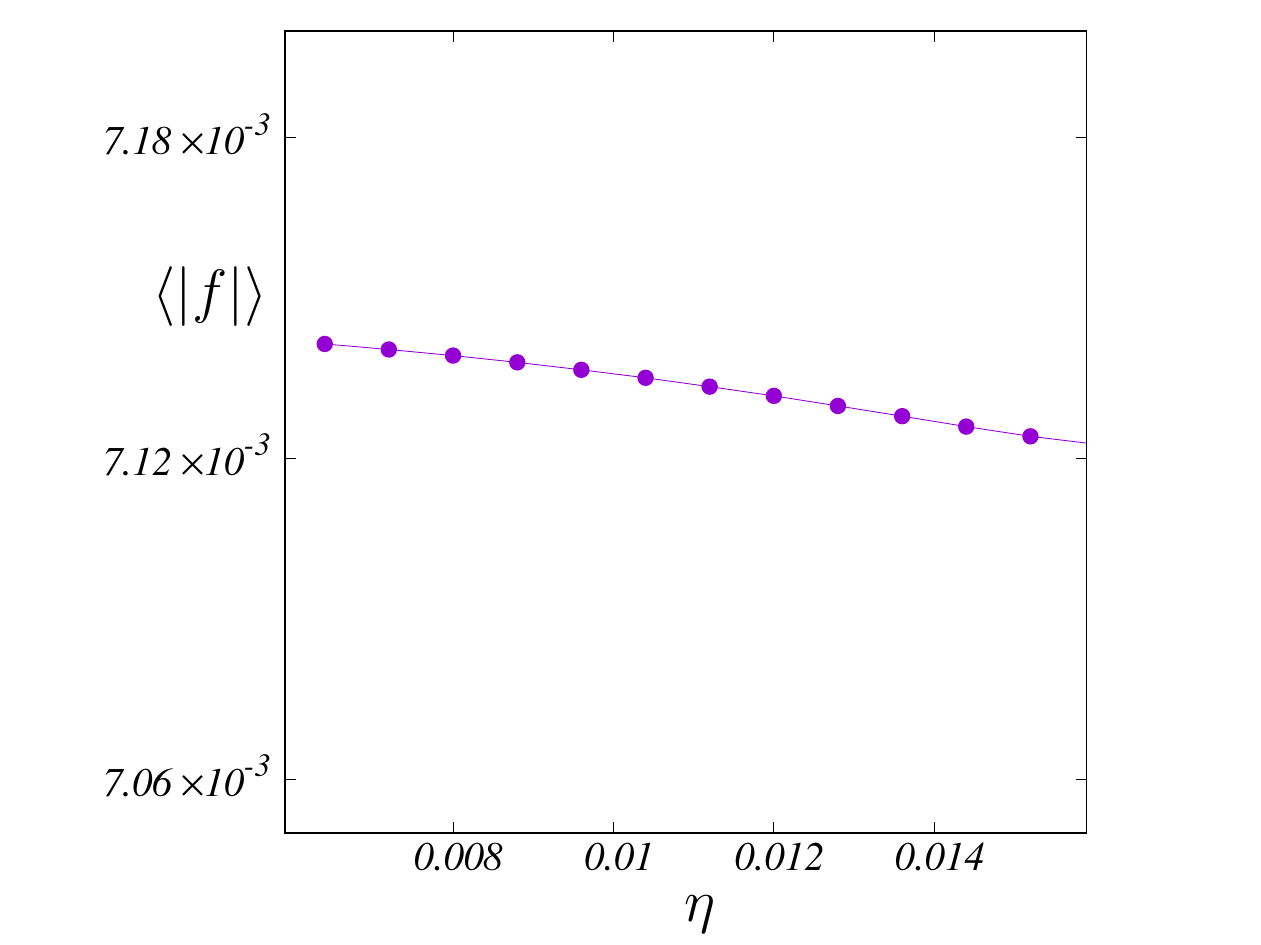}
\caption{The change in the mean value of the forces with increasing disorder strength for {\bf (a)} thermal and {\bf (b)} athermal systems. The scale has been magnified to show the effect of the small non-linear behaviour in the region of interest of our present study.}
\label{mean_force_fig}
\end{figure*}

\subsubsection{Thermal Disorder}
For small disorder, the distributions are symmetric about their average values, and the mean values of these fluctuations are $\langle \delta f_\parallel \rangle = 0$ and $\langle \delta f_\perp \rangle = 0$. This is true if one considers the linear response of the total energy of the system to the transformation $\delta f^{x(y)}_{ij} \to -\delta f^{x(y)}_{ij}$. This transformation leaves the Hamiltonian of the system invariant, to linear order. Similarly, the transformation leaves the Hamiltonian of the thermal force balanced crystal (Eq. (\ref{force_hamiltonian})) invariant. Hence at low temperatures, where the linear regime dominates, the mean value of the forces remains constant. In Fig. \ref{mean_force_fig} (a) we plot the variation of the mean value of the force in the thermal system obtained in our simulations, highlighting the almost constant behaviour as the temperature is varied.

\subsubsection{Athermal Disorder}

Within the linear theory, the mean value $f_0$ is independent of the polydispersity, as the mean positions of the particles are the crystalline positions of the original triangular lattice. However, there could be non-trivial corrections to this behaviour with increasing disorder. This is however a higher order effect, since in the linear theory the forces are completely determined by a {\it linear} ``Green's function" connecting the displacements to the disorder in the radii (Eq. (\ref{fourier_matrix_eq}) in the main text). Hence, for a given realization of the athermal disorder, the transformation $\{\eta_{i}\} \to \{-\eta_{i}\}$ transforms the force deviations as $\delta f^{x(y)}_{ij} \to -\delta f^{x(y)}_{ij}$. Hence to linear order, the force fluctuations are symmetric about their mean value. In Fig. \ref{mean_force_fig} (b) we plot the variation of the mean value of the force in the system obtained in our simulations, highlighting the small variation as the strength of the athermal disorder is increased.


\subsection{Fourier space representation}
Following the convention in Fig. \ref{lattice_figure}, the force balance on every grain $i$ can be expressed as
\begin{equation}
\vec{f}_{i0} + \vec{f}_{i1} + \vec{f}_{i2} + \vec{f}_{i3} + \vec{f}_{i4} + \vec{f}_{i5} = 0.
\end{equation}
Using the linearized expressions in Eq. (\ref{linearized_expansion}) this leads to equations relating the changes in the positions to the changes in the radii as
\begin{eqnarray}
\nonumber
\sum_{j = 0}^{5} C_{ij}^{xx}(\delta x_{i} - \delta x_{j}) + \sum_{j = 0}^{5} C_{ij}^{xy} (\delta y_{i} - \delta y_{j}) = -\sum_{j = 0}^{5}  C_{ij}^{x \sigma} (\delta \sigma_{i} + \delta \sigma_{j}),\\
\sum_{j = 0}^{5} C_{ij}^{yx}(\delta x_{i} - \delta x_{j}) + \sum_{j = 0}^{5}  C_{ij}^{yy} (\delta y_{i} - \delta y_{j}) = -\sum_{j = 0}^{5} C_{ij}^{y \sigma} (\delta \sigma_{i} + \delta \sigma_{j}).
\label{position1_eq}
\end{eqnarray}
Next, we define the Fourier transforms of the changes in positions and radii as
\begin{eqnarray}
\nonumber
\delta x (\vec{k})  = \sum_{\vec{r}} \exp(i \vec{k}. \vec{r}) \delta x (\vec{r}),\\
\nonumber
\delta y (\vec{k})  = \sum_{\vec{r}} \exp(i \vec{k}. \vec{r}) \delta y (\vec{r}),\\
\delta \sigma (\vec{k})  = \sum_{\vec{r}} \exp(i \vec{k}. \vec{r}) \delta \sigma (\vec{r}).
\label{fourier_transforms_eq}
\end{eqnarray}
Here $\vec{r} \equiv i$ label the sites of the triangular lattice whereas 
\begin{equation}
\vec{k} \equiv (k_x,k_y) \equiv \left( \frac{2 \pi l}{2 N}, \frac{2 \pi m}{N} \right),
\end{equation}
are the reciprocal lattice vectors of the triangular lattice \cite{horiguchi1972lattice}. 
Since the changes in the radii are i.i.d. variables, we have (using Eq. (\ref{change_radii_law}) in the main text)
\begin{equation}
\langle \delta \sigma (\vec{r})\delta \sigma (\vec{r'})  \rangle =  \eta^2 \sigma_0^2 \delta(\vec{r} - \vec{r}')  \int_{-1/2}^{1/2} d \xi ~\xi^2  =  \frac{\eta^2}{48} \delta(\vec{r} - \vec{r'}),
\label{real_sigma_eq}
\end{equation}
where we have used $\sigma_0 = 1/2$.
This can then be used to compute the correlations in Fourier space as
\begin{equation}
\langle \delta \sigma (\vec{k})  \delta \sigma (-\vec{k})  \rangle = \frac{\eta^2}{48}. 
\label{fourier_sigma_eq}
\end{equation}
It is also convenient to define the following Fourier coefficients
\begin{eqnarray}
\nonumber
\mathcal{F}_0(\vec{k}) &=& e^{-2 i k_x},\\
\nonumber
\mathcal{F}_1(\vec{k}) &=& e^{-i k_x-i k_y},\\
\nonumber
\mathcal{F}_2(\vec{k}) &=& e^{i k_x-i k_y},\\
\nonumber
\mathcal{F}_3(\vec{k}) &=& e^{2 i k_x},\\
\nonumber
\mathcal{F}_4(\vec{k}) &=& e^{i k_x+i k_y},\\
\mathcal{F}_5(\vec{k}) &=& e^{i k_y-i k_x}.
\end{eqnarray}
Next, multiplying Eq. (\ref{position1_eq}) by  $\exp(i \vec{k}. \vec{r})$ and summing over all sites $\vec{r} \equiv i$ leads to the following matrix equation at every $\vec{k}$
\begin{equation}
\left(
\begin{matrix} 
A^{xx}(\vec{k}) & A^{xy}(\vec{k}) \\
A^{yx}(\vec{k}) & A^{yy}(\vec{k})
\end{matrix}
\right)
\left(
\begin{matrix} 
\delta x (\vec{k}) \\
\delta y (\vec{k})
\end{matrix}
\right)
= 
\delta \sigma (\vec{k})
\left(
\begin{matrix} 
D^{x} (\vec{k}) \\
D^{y} (\vec{k})
\end{matrix}
\right),
\label{matrix_supp_eq}
\end{equation}
which is Eq. (\ref{fourier_matrix_eq}) in the main text. These matrix elements have the following explicit representations
 \begin{eqnarray}
\nonumber
A^{xx}(\vec{k}) &=& -\sum_{j=0}^{5} \mathcal{F}_j(\vec{k}) C^{xx}_{ij}(\phi) + \sum_{j=0}^{5} C^{xx}_{ij}(\phi),\\
\nonumber
A^{xy}(\vec{k}) &=& -\sum_{j=0}^{5} \mathcal{F}_j(\vec{k}) C^{xy}_{ij}(\phi) + \sum_{j=0}^{5} C^{xy}_{ij}(\phi) ,\\
\nonumber
A^{yx}(\vec{k}) &=& -\sum_{j=0}^{5} \mathcal{F}_j(\vec{k}) C^{yx}_{ij}(\phi) + \sum_{j=0}^{5} C^{yx}_{ij}(\phi),\\
A^{yy}(\vec{k}) &=& -\sum_{j=0}^{5} \mathcal{F}_j(\vec{k}) C^{yy}_{ij}(\phi) + \sum_{j=0}^{5} C^{yy}_{ij}(\phi).
\end{eqnarray} 
Similarly we have 
\begin{eqnarray}
\nonumber
D^{x} (\vec{k}) &=& -\sum_{j=0}^{5} \mathcal{F}_j(\vec{k}) C^{x\sigma}_{ij}(\phi) - \sum_{j=0}^{5}C^{x\sigma}_{ij}(\phi),\\
D^{y} (\vec{k}) &=& -\sum_{j=0}^{5} \mathcal{F}_j(\vec{k}) C^{y\sigma}_{ij}(\phi) - \sum_{j=0}^{5}C^{y\sigma}_{ij}(\phi).
\end{eqnarray}
Inverting Eq. (\ref{matrix_supp_eq}) leads to an expression for the Fourier transformed changes in positions in terms of the Fourier transformed changes in radii
 \begin{eqnarray}
\nonumber
\delta  x(\vec{k})= \alpha(\vec{k})\delta {\sigma}(\vec{k}),\\
\delta  y(\vec{k}) = \beta(\vec{k})\delta {\sigma}(\vec{k}),
\end{eqnarray}
which is Eq. (\ref{deltax_k_eq}) in the main text. 
Finally, an inverse Fourier transform and Eq. (\ref{fourier_sigma_eq}) yields the fluctuations in the positions at every site $i$
  \begin{eqnarray}
    \nonumber
\langle\delta x_{i}^2 \rangle=\frac{1}{2L^2} \left( \frac{\eta^2}{48} \right) \sum_{m=0}^{L-1}\sum_{l=0}^{2L-1}(\alpha(\vec{k})\alpha(-\vec{k})),\\
\nonumber
  \langle\delta y_{i}^2 \rangle=\frac{1}{2L^2} \left( \frac{\eta^2}{48} \right) \sum_{m=0}^{L-1}\sum_{l=0}^{2L-1}(\beta(\vec{k})\beta(-\vec{k})),\\
  \langle\delta x_{i} \delta y_{i}\rangle=\frac{1}{2L^2} \left( \frac{\eta^2}{48} \right) \sum_{m=0}^{L-1}\sum_{l=0}^{2L-1}(\alpha(\vec{k})\beta(-\vec{k})).
   \end{eqnarray} 
  Similarly, the fluctuations in the forces at every site $i$ can be computed by expressing the linearized expressions in Eq. (\ref{linearized_expansion}) in Fourier space. We have (with $j = 0$ to $5$, see Fig. \ref{lattice_figure})
   \begin{eqnarray}
   \nonumber
   \langle\delta f^x_{ij} \delta f^x_{ij}\rangle &=& \frac{1}{2L^2} \left( \frac{\eta^2}{48} \right) \sum_{m=0}^{L-1}\sum_{l=0}^{2L-1} [C_{ij}^{xx}(1 - \mathcal{F}_j(\vec{k}))\alpha(\vec{k})+C_{ij}^{xy}(1 - \mathcal{F}_j(\vec{k}))\beta(\vec{k})+C_{ij}^{x\sigma}(1 + \mathcal{F}_j(\vec{k}))]\\
&&   \times [C_{ij}^{xx}(1 - \mathcal{F}_j(\vec{k})^{-1})\alpha(-\vec{k})+C_{ij}^{xy}(1 - \mathcal{F}_j(\vec{k})^{-1})\beta(-\vec{k})+C_{ij}^{x\sigma}(1 + \mathcal{F}_j(\vec{k})^{-1})],
\label{force_correlator1_eq}
 \end{eqnarray} 
 \begin{eqnarray}
\nonumber
\langle\delta f^y_{ij} \delta f^y_{ij}\rangle &=& \frac{1}{2L^2} \left( \frac{\eta^2}{48} \right) \sum_{m=0}^{L-1}\sum_{l=0}^{2L-1} [C_{ij}^{yx}(1 - \mathcal{F}_j(\vec{k}))\alpha(\vec{k})+C_{ij}^{yy}(1 - \mathcal{F}_j(\vec{k}))\beta(\vec{k})+C_{ij}^{y\sigma}(1 + \mathcal{F}_j(\vec{k}))]\\
&&   \times [C_{ij}^{yx}(1 - \mathcal{F}_j(\vec{k})^{-1})\alpha(-\vec{k})+C_{ij}^{yy}(1 - \mathcal{F}_j(\vec{k})^{-1})\beta(-\vec{k})+C_{ij}^{y\sigma}(1 + \mathcal{F}_j(\vec{k})^{-1})],
\label{force_correlator2_eq}
\end{eqnarray} 
\begin{eqnarray}
\nonumber
\langle\delta f^x_{ij} \delta f^y_{ij}\rangle &=& \frac{1}{2L^2} \left( \frac{\eta^2}{48} \right) \sum_{m=0}^{L-1}\sum_{l=0}^{2L-1} [C_{ij}^{xx}(1 - \mathcal{F}_j(\vec{k}))\alpha(\vec{k})+C_{ij}^{xy}(1 - \mathcal{F}_j(\vec{k}))\beta(\vec{k})+C_{ij}^{x\sigma}(1 + \mathcal{F}_j(\vec{k}))]\\
&&  \times  [C_{ij}^{yx}(1 - \mathcal{F}_j(\vec{k})^{-1})\alpha(-\vec{k})+C_{ij}^{yy}(1 - \mathcal{F}_j(\vec{k})^{-1})\beta(-\vec{k})+C_{ij}^{y\sigma}(1 + \mathcal{F}_j(\vec{k})^{-1})].
\label{force_correlator3_eq}
\end{eqnarray}  
In Fig. \ref{variance_figure} we plot the variance in the components of the forces computed from numerical simulations at different polydispersities and packing fractions along with the above theoretical predictions. We find that the predictions from this theory match the simulations exactly at low $\eta$ and begin to deviate at higher values of $\eta$ where the higher order terms in the perturbation expansion begin to play a role.

Finally, we can also use the formalism developed here to compute the fluctuations in the bond angles $\sin\delta\theta_{ij}$ in the system. Once again expanding to linear order about the crystalline values, we have
 \begin{equation}
 \sin \delta\theta_{ij} = B_{ij}^{x}\delta x_{ij} + B_{ij}^{y}\delta y_{ij},
 \end{equation}
 where the coefficients are given by
\begin{eqnarray}
\nonumber
 B_{ij}^{x} &=& -\frac{\sin\theta_{ij}^{0}}{\sqrt{\phi_c/\phi}},\\
 B_{ij}^{y} &=& \frac{\cos\theta_{ij}^{0}}{\sqrt{\phi_c/\phi}}.
\end{eqnarray}  
 We then have
 \begin{eqnarray}
 \nonumber
\langle (\sin\delta\theta_{ij})^2\rangle &= &\frac{1}{2L^2}\left(\frac{\eta^2}{48}\right)\sum_{m=0}^{L-1}\sum_{l=0}^{2L-1}[B_{ij}^{x}(1 - \mathcal{F}_j(\vec{k}))\alpha(\vec{k})+B_{ij}^{y}(1 - \mathcal{F}_j(\vec{k}))\beta(\vec{k})]\\
 && \times [B_{ij}^{x}(1 - \mathcal{F}_j(\vec{k})^{-1})\alpha(-\vec{k})+B_{ij}^{y}(1 - \mathcal{F}_j(\vec{k})^{-1})\beta(-\vec{k})].
 \label{sin_deltheta_variance_eq}
\end{eqnarray}
The above expression along with the variance in the forces can then be used to compute the distribution of the orthogonal components of the forces $p(f_{\perp})$ as described in the main text.

\begin{figure}[t!]
\includegraphics[scale=0.5]{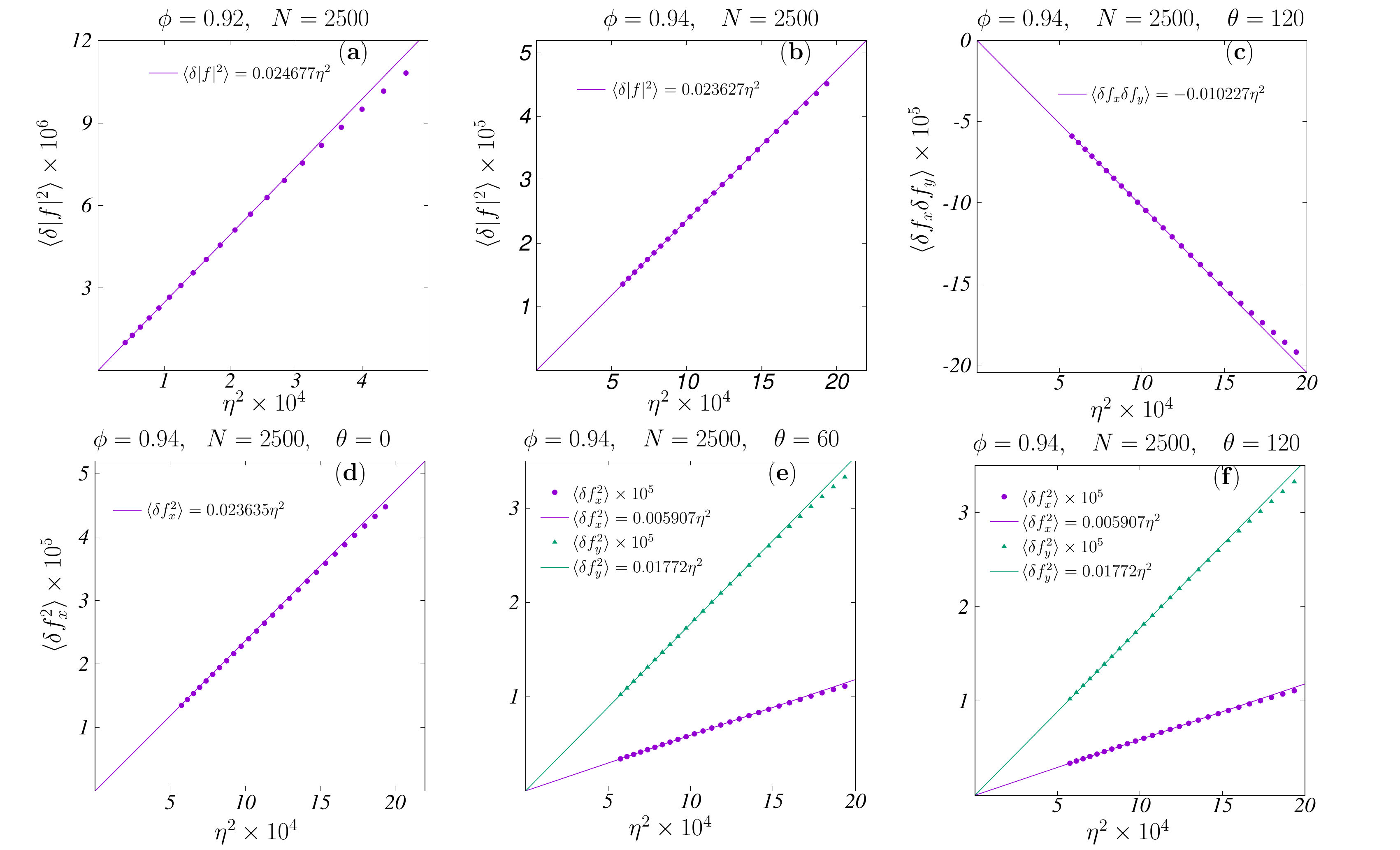}
\centering
\caption{Plot of the variance in the components of the forces computed from numerical simulations at different polydispersities ($\eta$) and packing fractions ($\phi$) along with the theoretical predictions in Eqs. (\ref{force_correlator1_eq}), (\ref{force_correlator2_eq}) and (\ref{force_correlator3_eq}). {\bf (a)} Variance in the change in magnitude of the forces $\langle\delta |f|^2 \rangle$ for $\phi = 0.92$. {\bf (b)} $\langle\delta |f|^2 \rangle$ for $\phi = 0.94$. These correlations have been computed by averaging over all six directions of the lattice ($j = 0$ to $5$). {\bf (c)} $\langle\delta f_x \delta f_y \rangle$ for $\phi = 0.94$ at an angle $\theta = 120 ~(j = 2)$. {\bf (d)} $\langle\delta f_x \delta f_x\rangle$ for $\phi = 0.94$ at $\theta = 0 ~(j = 0)$. {\bf (e)} $\langle\delta f_x \delta f_x\rangle$ and $\langle\delta f_y \delta f_y\rangle$ for $\phi = 0.94$,  and $\theta = 60 ~(j =1)$. {\bf (f)} $\langle\delta f_x^2 \rangle$ and $\langle\delta f_y^2 \rangle$ for $\phi = 0.94$ at $\theta = 120 ~(j =2)$. All quantities displayed have been computed for a system size $N =2500$. We find that the predictions from the theory match the simulations exactly at low $\eta$ and begin to deviate at higher values of $\eta$ where the higher order terms in the perturbation expansion begin to play a role. }
\label{variance_figure}
\end{figure}


\subsection{Exact Series Expression for $p(f_{\perp})$}

Let $x$ and $y$ be two uncorrelated random variables 
The probability distribution of the variable $z = xy$ can then be expressed as
\begin{equation}
p(z) = \int_{-\infty}^{\infty} \int_{-\infty}^{\infty}p(x)p(y)\delta(z-xy) dx dy =  \int_{-\infty}^{\infty} p(x) p\left(\frac{z}{x}\right)\frac{1}{|x|}dx.
\label{product_distribution_eq}
\end{equation}   
Furthermore suppose $x$ and $y$ are normally distributed
with means $\mu_1$ and $\mu_2$ and standard deviations $\sigma_1$ and $\sigma_2$ respectively. i.e.
\begin{equation}
p(x) = \frac{1}{\sqrt{2\pi\sigma_1^2}}e^{-\frac{(x-\mu_1)^2}{2 \sigma_1^2}},
\end{equation}
and
\begin{equation}
p(y) = \frac{1}{\sqrt{2\pi\sigma_2^2}}e^{-\frac{(y-\mu_2)^2}{2 \sigma_2^2}}.
\end{equation}
For such variables with non-zero means ($\mu_1 \ne 0, \mu_2 \ne 0$), the integral in Eq. (\ref{product_distribution_eq}) is non-trivial. However, it is still possible to obtain an exact series representation. The final expression of this integral has the following form \cite{cui2016exact}
\begin{small}
 \begin{equation}
 p(z) = e^{-\left(\frac{\mu_1^2}{2\sigma_1^2} + \frac{\mu_2^2}{2\sigma_2^2}\right)} \sum_{n=0}^{\infty}\sum_{m=0}^{2n}\frac{z^{2n-m}|z|^{m-n}\sigma_1^{m-n-1}}{\pi (2n)!\sigma_2^{m-n+1}}{2n\choose m}\left(\frac{\mu_1}{\sigma_1^2}\right)^m \left(\frac{\mu_2}{\sigma_2^2}\right)^{2n-m}K_{m-n}\left(\frac{|z|}{\sigma_1\sigma_2}\right).
 \end{equation}
 \end{small}
Here $K_{m-n}(.)$ is the modified Bessel function of the second kind of order $(m-n)$. This function displays non-analytic behaviour at $z=0$.
\begin{figure*}[t!]
\centering
\includegraphics[scale=0.55]{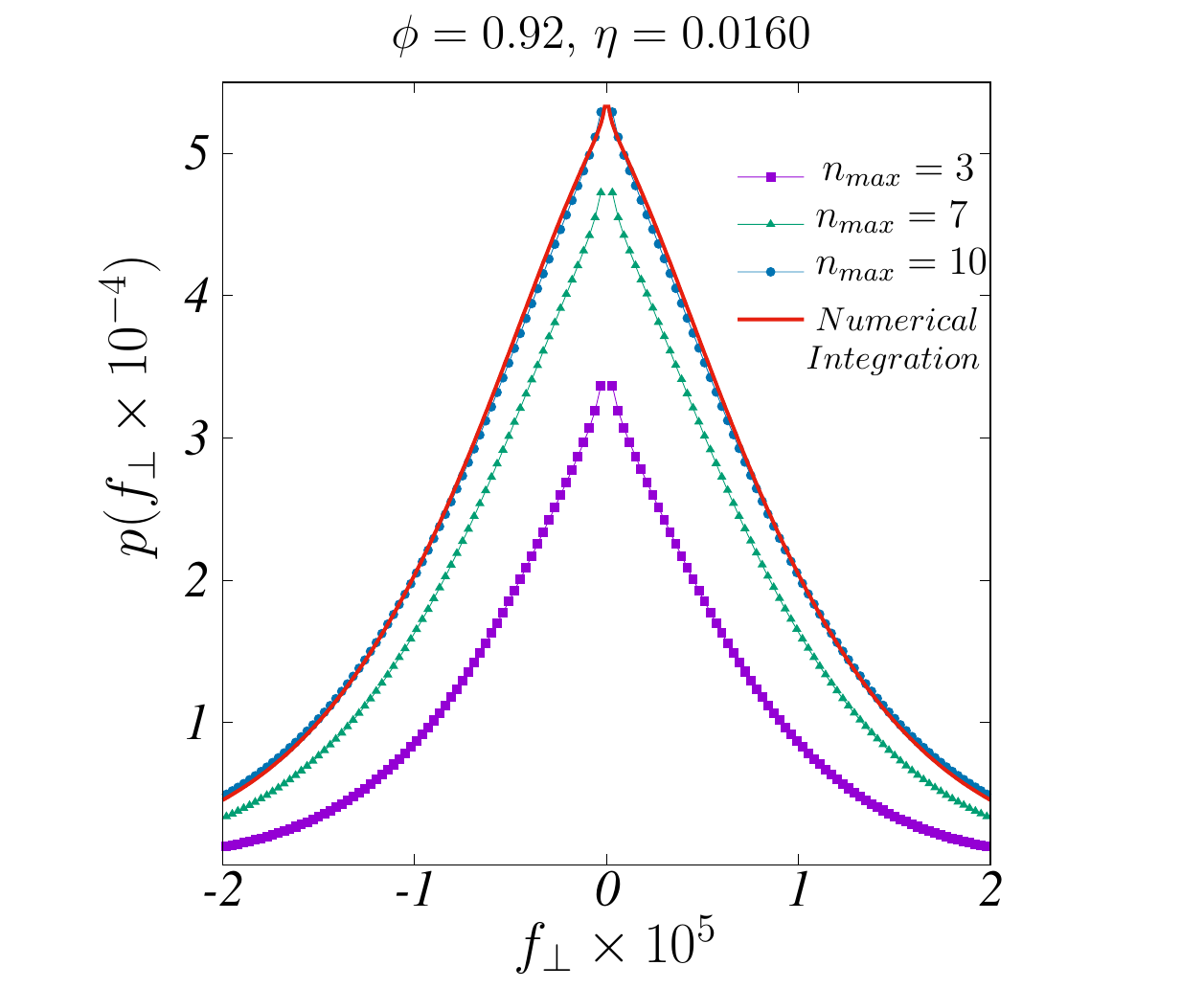}
\caption{Plot of the distribution of $f_\perp$ obtained from the exact series expression in Eq. (\ref{simplified_expression_eq}) (with increasing number of terms $n_{max}$), along with a direct integration of the expression in Eq. (\ref{product_distribution_eq}). The two expressions converge as $n_{max}$ is increased.}
\label{bessel_expression_fig}
\end{figure*}
The case we are considering $f_{\perp} = |f| \sin(\delta \theta)$ has $\langle \sin(\delta \theta) \rangle = 0$, and $\langle |f| \rangle = f_0 \ne 0$. The final expression of this integral can be simplified to the following form
 \begin{equation}
 p(f_{\perp}) = e^{-\left(\frac{f_0^2}{2\sigma_1^2}\right)} \sum_{n=0}^{\infty}\frac{|f_{\perp}|^{n}\sigma_1^{n-1}}{\pi (2n)!\sigma_2^{n+1}}\left(\frac{f_0}{\sigma_1^2}\right)^{2n} K_{n}\left(\frac{|f_{\perp}|}{\sigma_1\sigma_2}\right),
 \label{simplified_expression_eq}
 \end{equation}
 where $\sigma_1$ and $\sigma_2$ are the standard deviations of the fluctuations in the forces $|f|$ and $\sin(\delta \theta)$ respectively, which we have computed in Eqs. (\ref{force_correlator1_eq}~--~\ref{force_correlator3_eq}) and Eq. (\ref{sin_deltheta_variance_eq}).
In Fig. \ref{bessel_expression_fig} we plot the distribution of $f_\perp$ obtained from the above series (with an increasing number of terms), and a direct integration of the expression in Eq. (\ref{product_distribution_eq}), showing the convergence of the above exact series expression to the numerically integrated curve displayed in Fig. \ref{fy_figure} in the main text.

\subsection{Joint Distribution of $|f|$ and $\sin(\delta \theta)$}

In this Section we analyze the correlations between the variables $|f|$ and $\sin(\delta \theta)$ which we use to compute the distribution of $f_{\perp} = |f| \sin(\delta \theta)$.

As we show in the main text, the force balance conditions on every particle yield $2 N$ equations for the  $2 N$ position variables $\{x_i, y_i\}$.
Since this system of equations is invertible, these position variables are linearly independent. 
The forces, and the relative bond angles then be derived from these positions by linearizing the force law.
As the forces are derived from the bond distances, not all the forces in the system are independent. In the triangular lattice arrangement, there are $N_C = 6 N_G$ {\it vector} bond variables ($f^x_{ij}$ and $f^y_{ij}$), where $N_C$ is the total number of contacts, and $N_G$ is the number of particles in the system.  However, since $\vec{f}_{ij} = -\vec{f}_{ji}$, these reduce to $3 N_G$ vector variables. Clearly, the representation of the degrees of freedom in the system in terms of the forces then is an overparametrization. There are therefore additional constraints that these variables must satisfy. It is easy to see that these are the loop constraints providing $2 N_G$ vector equations, leaving the system with $N_G$ independent vector variables.

In the derivation of the distribution of $f_{\perp}$, we have assumed that the parametrization of the system in terms of the magnitude of each force $|f_{ij}|$ and the relative angles measured in terms of the original lattice directions $\delta \theta_{ij} = \theta_{ij} - \theta_{ij}^0$. This is in effect a $\{f^x_{ij}, f^y_{ij} \} \to \{|f_{ij}|,\delta \theta_{ij} \} $ transformation. Therefore the loop constraints still need to be imposed on these variables. However, these are higher order correlations as we show below. In our linear theory, we can compute the correlations in these variables to leading order exactly. We find
\begin{eqnarray}
\nonumber
\langle |f|^2 \rangle = 0.0241 \eta^2,\\
\langle \sin^2 \delta \theta \rangle = 6.62 \times 10^{-3} \eta^2.
\end{eqnarray}
However, as our linear theory predicts that the correlation $\langle |f| \sin \delta \theta \rangle$ is exactly zero to lowest order, we have also measured the following correlation in our simulations and find 
\begin{eqnarray}
|\langle |f| \sin \delta \theta \rangle| = 1.9 \times 10^{-10} \eta^2.
\end{eqnarray}
Therefore, to leading order the cross-correlations between these variables is very small in comparison to their individual fluctuations, justifying our uncorrelated computation. In Fig. \ref{force_variance_fig} we plot the variance in the components of the forces, as well as the above correlations. Taking the uncorrelated assumption further, the joint distribution of the variables $|f| $ and $\sin \delta \theta$ can be written as a product form
\begin{equation}
p(|f|,\sin \delta \theta) = p(|f|) p(\sin \delta \theta).
\end{equation}
In Fig. \ref{joint_distribution_fig} we plot the joint distribution of the variables $|f|$ and $\sin \delta \theta$ obtained from simulations, as well as from the above uncorrelated product form, showing that to leading order this distribution can be reproduced using the marginal distributions of each of these variables.

\begin{figure*}[]
\centering
\includegraphics[scale=0.45]{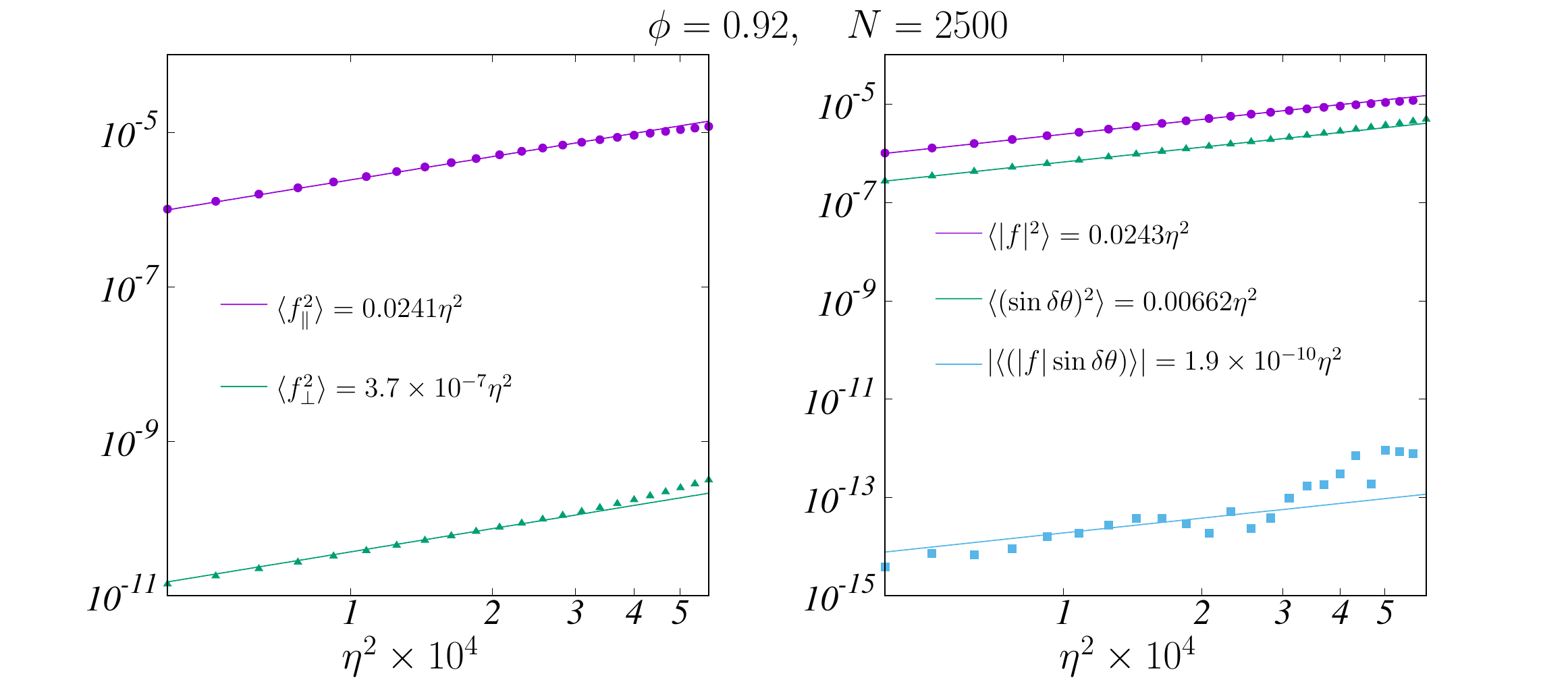}
\caption{(Left) Plot of the variance in the components of the forces, showing that the fluctuations in $f_{\perp}$ are much smaller than the fluctuations in  $f_{||}$. (Right) The correlation between $|f_{ij}|$ and $\sin \delta \theta_{ij}$ is much smaller in comparision to their individual fluctuations.}
\label{force_variance_fig}
\end{figure*}

\begin{figure}[t!]
\centering
\includegraphics[scale=0.45]{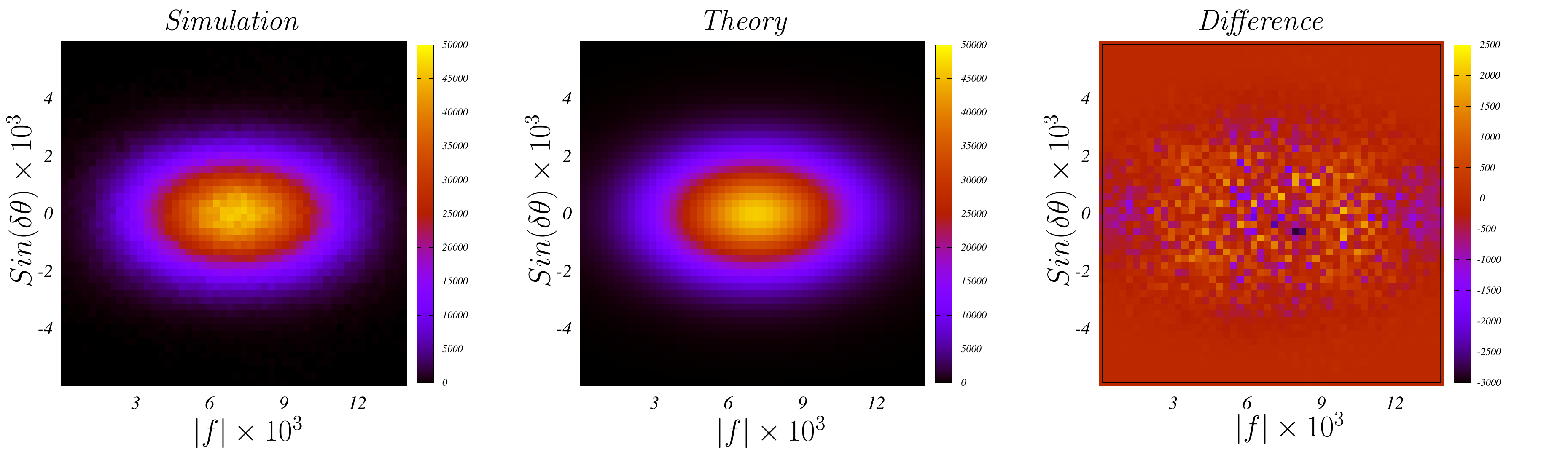}
\caption{The joint distribution of the variables $|f|$ and $\sin \delta \theta$, observed in the numerics (left) and using an uncorrelated form predicted from the linear theory (middle). The (right) panel displays the difference between these two distributions, showing that to leading order this distribution can be reproduced using the marginal distributions of each of these variables.}
\label{joint_distribution_fig}
\end{figure}

\subsection{Distribution of overlap lengths}

\begin{figure}[t!]
\includegraphics[scale=0.8]{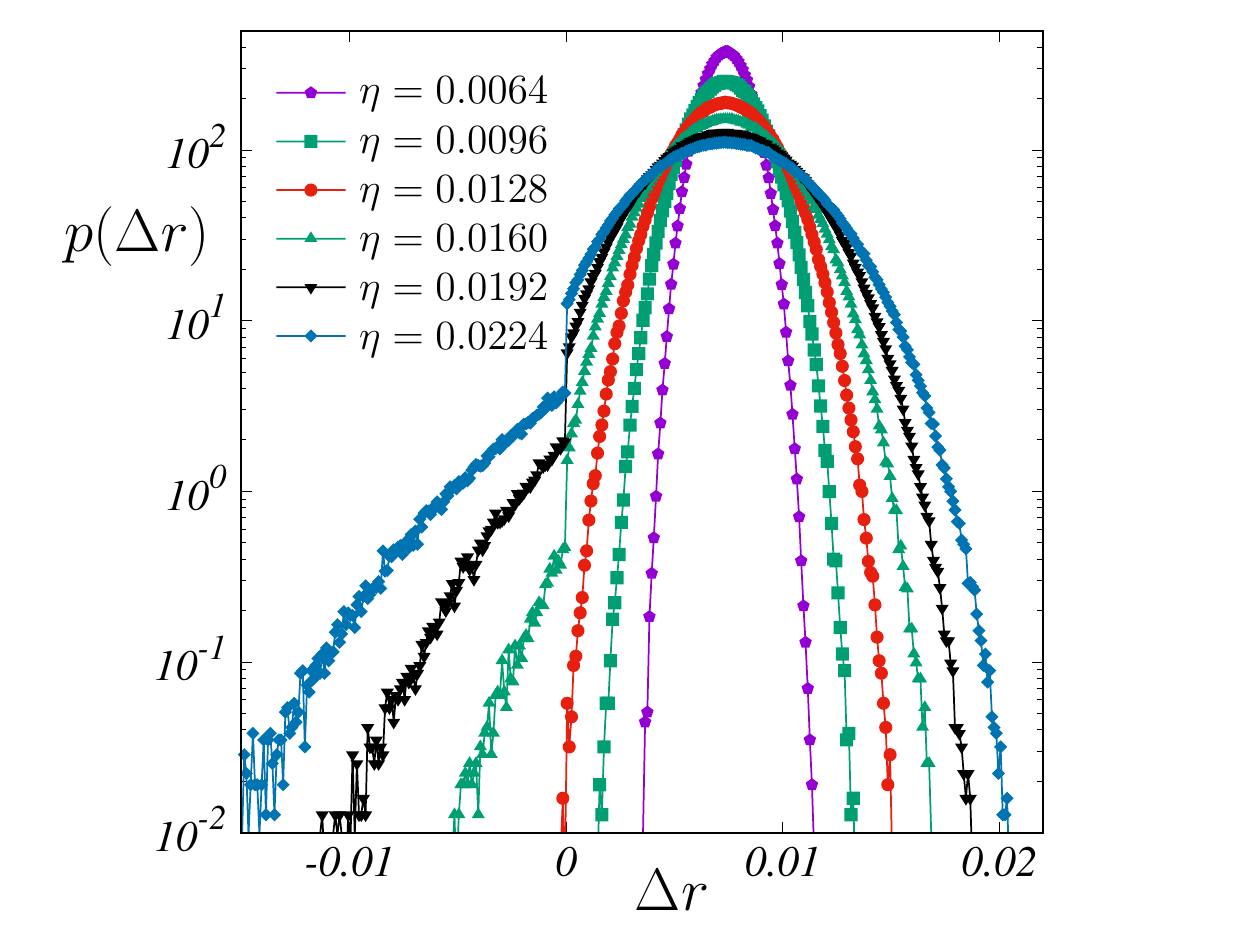}
\centering
\caption{Distribution of overlap lengths $\Delta r_{ij} = \sigma_{i}+\sigma_{j}-|\vec{r}_i - \vec{r}_{j}|$ between particles at different polydispersities ($\eta$) for $\phi = 0.92$. For low polydispersities all overlaps are positive, i.e. there are no broken contacts. At higher $\eta$, contacts break and the overlap distribution develops a discontinuity, signifying system spanning rearrangements.}
\label{overlap_figure}
\end{figure}

Finally, we investigate the origin of the small deviation in the average coordination predicted by the theory and those obtained from numerical simulations as shown in Fig. \ref{zz_figure} in the main text. 
To study the process of contact breaking in the system, we analyze the distribution of overlap lengths $\Delta r_{ij} = \sigma_{i}+\sigma_{j}-|\vec{r}_i - \vec{r}_{j}|$ between neighbouring particles $i$ and $j$ in the system. $\Delta r > 0$ represent real (force bearing) contacts, whereas $\Delta r < 0$ represent the broken contacts in the system. Since we have focused on harmonic interactions in this study, the distribution $p(\Delta r)$ for $\Delta r > 0$ is exactly the distribution of forces $p(|f|)$ (with a suitable normalization). As the crystalline systems we study have $\Delta \phi >0$, all the contacts bear a finite force and $\Delta r > 0$ at $\eta = 0$, with $p(\Delta r) = \delta(\Delta r - f_0)$ (as given in Eq. (\ref{overlap_eq})). With increasing polydispersity, this distribution broadens and contacts begin to break, populating the $\Delta r < 0$ regions. We plot this distribution for different polydispersities in Fig. \ref{overlap_figure}. Surprisingly, although both regions are well fit by Gaussians, they are separated by a discontinuity. This suggests that as a bond between two particles breaks, $\Delta r$ moves a finite distance away from $0$. We have tested that this is indeed the case by gradually increasing the polydispersity and following the evolution of the broken contacts in the system. We attribute this ``kick" felt by these bonds as originating from the system spanning rearrangements that occur in response to a contact breaking event.
As our prediction for $z$ in Eq. (\ref{coordination_eq}) in the main text was obtained from the distribution of the force magnitudes extrapolated to the unphysical regions $|f|<0$, the finite discontinuity in the $\Delta r$ distribution explains the origin of the shift in the numerically obtained $z$ and those predicted by the theory. It would be interesting to extend our methods to develop an explanation for this non-linear contact breaking process.

\clearpage
\end{appendix}

\end{widetext}

\end{document}